\documentclass[aps,prl,twocolumn,showpacs,amsmath,amssymb,superscriptaddress,reprint]{revtex4-1}

\usepackage[utf8]{inputenc}
\usepackage{graphicx}
\usepackage[colorlinks=true,citecolor=blue]{hyperref}
\usepackage{units}
\usepackage{verbatim} 

\renewcommand{\vec}[1]{\mathbf{#1}}
\newcommand{\kBT}{k_\text{B}T}

\def\kB {k_\text{B}}

\begin{document}
\title{Optimal Quantum Interference Thermoelectric Heat Engine with Edge States}
\author{Peter Samuelsson}
\affiliation{Physics Department and NanoLund, Lund University, Box 118, SE-22100 Lund, Sweden}
\author{Sara Kheradsoud}
\affiliation{Physics department and NanoLund, Lund University, Box 118, SE-22100 Lund, Sweden}
\author{Bj\"orn Sothmann}
\affiliation{Theoretische Physik, Universität Duisburg-Essen and CENIDE, D-47048 Duisburg, Germany}

\date{\today}

\begin{abstract}
We show theoretically that a thermoelectric heat engine, operating exclusively due to quantum-mechanical interference, can reach optimal linear-response performance. A chiral edge state implementation of a close-to-optimal heat engine is proposed in an electronic Mach-Zehnder interferometer with a mesoscopic capacitor coupled to one arm.  We demonstrate that the maximum power and corresponding efficiency can reach 90\% and 83\%, respectively, of the theoretical maximum. The proposed heat engine can be realized with existing experimental techniques and has a performance robust against moderate dephasing.
\end{abstract}

\maketitle

\paragraph{Introduction.--}
Thermoelectric effects have received a considerable amount of interest during the past years~\cite{shakouri_recent_2011}. They allow for the conversion of waste heat back into useful electricity which can be of use for energy harvesting applications. A special emphasis has been laid on thermoelectric effects in mesoscopic and nanoscale conductors, both from a theoretical~\cite{streda_quantised_1989,beenakker_theory_1992,hicks_effect_1993,hicks_thermoelectric_1993,nakpathomkun_thermoelectric_2010,karlstrom_increasing_2011} as well as from an experimental perspective~\cite{molenkamp_quantum_1990,staring_coulomb-blockade_1993,scheibner_thermopower_2005,svensson_lineshape_2012,svensson_nonlinear_2013}. The discrete nature of electronic states in such conductors turns them into good energy filters which consecutively can give rise to a high thermoelectric efficiency~\cite{mahan_best_1996}. Recently, multiterminal heat engines, which allow for the spatial separation of heat flow and electric power generation, have been proposed~\cite{entin-wohlman_three-terminal_2010,sanchez_optimal_2011,sothmann_rectification_2012,sothmann_magnon-driven_2012,jiang_three-terminal_2013,machon_nonlocal_2013,bergenfeldt_hybrid_2014,mazza_thermoelectric_2014,mazza_separation_2015,sothmann_thermoelectric_2015} and also realized experimentally~\cite{hartmann_voltage_2015,roche_harvesting_2015,thierschmann_three-terminal_2015}.   

Heat engines utilizing quantum-mechanical coherence, a fundamental property of mesoscopic and nanoscale transport, however, have received limited attention \cite{scully_extracting_2003,rosnagel_nanoscale_2014}. Arguably, quantum coherence is most strikingly manifested via interference effects. As the most prominent example, electronic interferometers based on chiral transport in quantum Hall edge channels can display visibilities of coherent oscillations up to $90\%$~\cite{neder_interference_2007}. Thermoelectric heat engines with edge states have been investigated theoretically~\cite{sothmann_quantum_2014,lopez_thermoelectric_2014,sanchez_chiral_2015,sanchez_heat_2015,sanchez_effect_2016,hofer_quantum_2015}, largely motivated by the perspective of heat-charge separation and performance enhancement~\cite{benenti_thermodynamic_2011,saito_thermopower_2011,sanchez_thermoelectric_2011,brandner_strong_2013,brandner_multi-terminal_2013} due to the broken time-reversal symmetry in the quantum Hall regime. Thermal transport properties
have also been investigated experimentally~\cite{granger_observation_2009,altimiras_non-equilibrium_2010,nam_thermoelectric_2013,jezouin_quantum_2013} and theoretically~\cite{levkivskyi_energy_2012,aita_heat_2013,vannucci_interference-induced_2015,goldstein_suppression_2016}. Recently, an edge state heat engine based on an electronic Mach-Zehnder interferometer was proposed~\cite{hofer_quantum_2015}, predicting large efficiency and power output, comparable to nanoscale heat engines based on, e.g., quantum dots~\cite{nakpathomkun_thermoelectric_2010,jordan_powerful_2013}.

\begin{figure}
	\includegraphics[width=.17\textwidth]{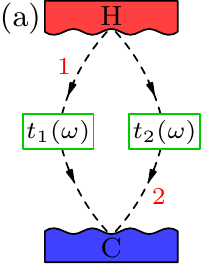}
	\includegraphics[width=.3\textwidth]{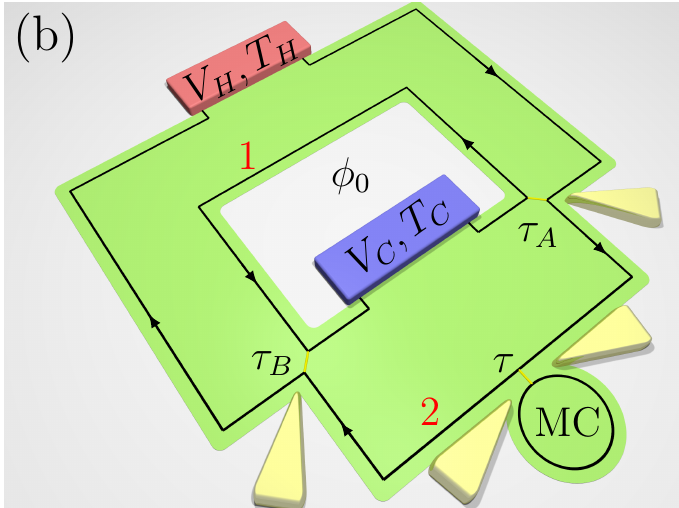}
	\caption{\label{fig:setup} (a) Schematic of a thermoelectric two-path interferometer. Each path $i=1,2$ from the hot (H) to the cold (C) terminal contains a phase coherent scatterer with energy-dependent transmission amplitude $t_i(\omega)$. (b) Realization of (a) in a two-terminal Mach-Zehnder interferometer implemented with edge states. The active edge state is denoted with a solid line, chirality shown with arrows. A mesoscopic capacitor (MC) is coupled to one interferometer arm via a quantum point contact with transparency $\tau$. Two additional quantum point contacts, transparencies $\tau_A,\tau_B$, form beam splitters. The hot (cold) terminal is kept at temperature $T_H$ ($T_C$) and potential $V_H$ ($V_C$). By tuning $\tau$ and the path phase difference $\phi_0$, the interferometer transmission probability can become steplike in energy. Close-to-optimal linear-response thermoelectric performance is obtained by adjusting step height and position, via $\tau_A,\tau_B$ and the capacitor resonance energy.}
\end{figure}

However, two still unanswered, fundamentally important questions are the following: what is the optimal  thermoelectric performance of a system based only on interference, and can such an optimal system be realized with edge states? In this Letter we show that a quantum interference thermoelectric heat engine can reach the theoretically maximal~\cite{whitney_most_2014,whitney_finding_2015}, single mode, power production and corresponding efficiency at linear response. Moreover, we propose an edge state realization based on a two-terminal Mach-Zehnder interferometer with a mesoscopic capacitor coupled to one interferometer arm (Fig.~\ref{fig:setup}), with a close-to-optimal performance. Our experimentally feasible proposal demonstrates that quantum interference can be harnessed for a dramatic performance boost of nanoscale and mesoscopic heat engines.

\paragraph{Quantum interference thermoelectrics.--}
We first present a compelling, physically transparent analysis of a generic, noninteracting two-path thermoelectric interferometer, a careful scattering theory investigation of the thermoelectric heat engine performance in the linear-response regime follows below. A schematic of the two-terminal interferometer is shown in Fig.~\ref{fig:setup}: an electron emitted from the hot reservoir (H) can take two different paths, $i=1$ or $2$, to the cold reservoir (C). Each path contains a phase-coherent scatterer with transmission amplitude $t_i(\omega)$. The total amplitude to propagate from H to C is then $A(\omega) \propto t_1(\omega)+e^{i\phi_0}t_2(\omega)$ where the energy-independent path phase difference $\phi_0$ accounts for, e.g., an enclosed Aharonov-Bohm flux.

To investigate the thermoelectric properties due to interference only, we require the two scatters to have {\it no individual thermoelectric response}, i.e., zero Seebeck and Peltier effects. This amounts to energy independent transmission probabilities $T_i=|t_i(\omega)|^2$, i.e., amplitudes of the form $t_i(\omega)=\sqrt{T_i}e^{i\alpha_i(\omega)}$ where the energy dependence is restricted to the phases $\alpha_i(\omega)$. The total transmission probability $|A(\omega)|^2$ can then be written
\begin{equation}
|A(\omega)|^2\propto T_1+T_2+2\sqrt{T_1T_2}\cos[\alpha_1(\omega)-\alpha_2(\omega)-\phi_0].
\label{prob}
\end{equation} 
Given Eq.~\eqref{prob}, how good can the thermoelectric performance of this, effectively single mode, interferometer be? In particular, of key importance for operation as a heat engine, what is the maximum electrical power produced and the corresponding heat-to-work conversion efficiency? For these quantities, it is known~\cite{whitney_most_2014,whitney_finding_2015} that the optimal thermoelectric should have a step-function-in-energy transmission probability, i.e., a sharp rise (or drop), from zero to maximum, over a narrow energy interval (compared to the background temperature). This is fulfilled for the probability $|A(\omega)|^2$ in Eq.~\eqref{prob} if and only if i) the total phase $\alpha_1(\omega)-\alpha_2(\omega)-\phi_0$ changes abruptly, as a function of energy, from $0$ to $\pi$ (or $\pi$ to $0$) and ii) the transmission probabilities of the two scatterers are equal, $T_1=T_2$.

\paragraph{System and transmission probability.--}
To answer the question how such an optimal interferometer could be realized experimentally, we consider a two-terminal electronic Mach-Zehnder interferometer implemented with edge states in a conductor in the integer quantum Hall regime, see Fig.~\ref{fig:setup}. The edges of the interferometer arms $1$ and $2$ have lengths $L_1$ and $L_2$ respectively. Arm $2$ contains a mesoscopic capacitor~\cite{buttiker_mesoscopic_1993,gabelli_violation_2006,feve_-demand_2007}, a small loop of length $L$, coupled to the edge via a quantum point contact with transparency $1-\tau$. The capacitor effectively acts as a quantum dot, with a level spacing $\Delta=2\pi \hbar v_\text{D}/L$, where $v_\text{D}$ is the edge state drift velocity, coupled to the interferometer arm. Two additional quantum point contacts constitute beam splitters with transparencies $\tau_A$ and $\tau_B$ respectively. All transparencies $\tau, \tau_A$ and $\tau_B$ are independent of energy and can be tuned electrostatically between $0$ and $1$. For simplicity (and in contrast to Ref.~\cite{hofer_quantum_2015}) we assume that the total path length difference is negligible, i.e. $[L_1-(L_2+L)]k_{\mathrm{B}}T/(\hbar v_\text{D}) \ll 1$ where $T$ is the background temperature, the effect of a finite path length difference is treated below. We stress that the path lengths can be controlled by additional electrostatic gates~\cite{ji_electronic_2003}. Moreover, for a top gate controlling the capacitor potential charging effects are suppressed~\cite{feve_-demand_2007}, motivating our noninteracting approach.
We remark that going beyond linear response, electron-electron interactions in the interferometer need to be accounted for; see, e.g., a recent investigation of the visibility of Mach-Zehnder interferometers~\cite{ngo_dinh_analytically_2013}.

\begin{figure}
	\includegraphics[width=\columnwidth]{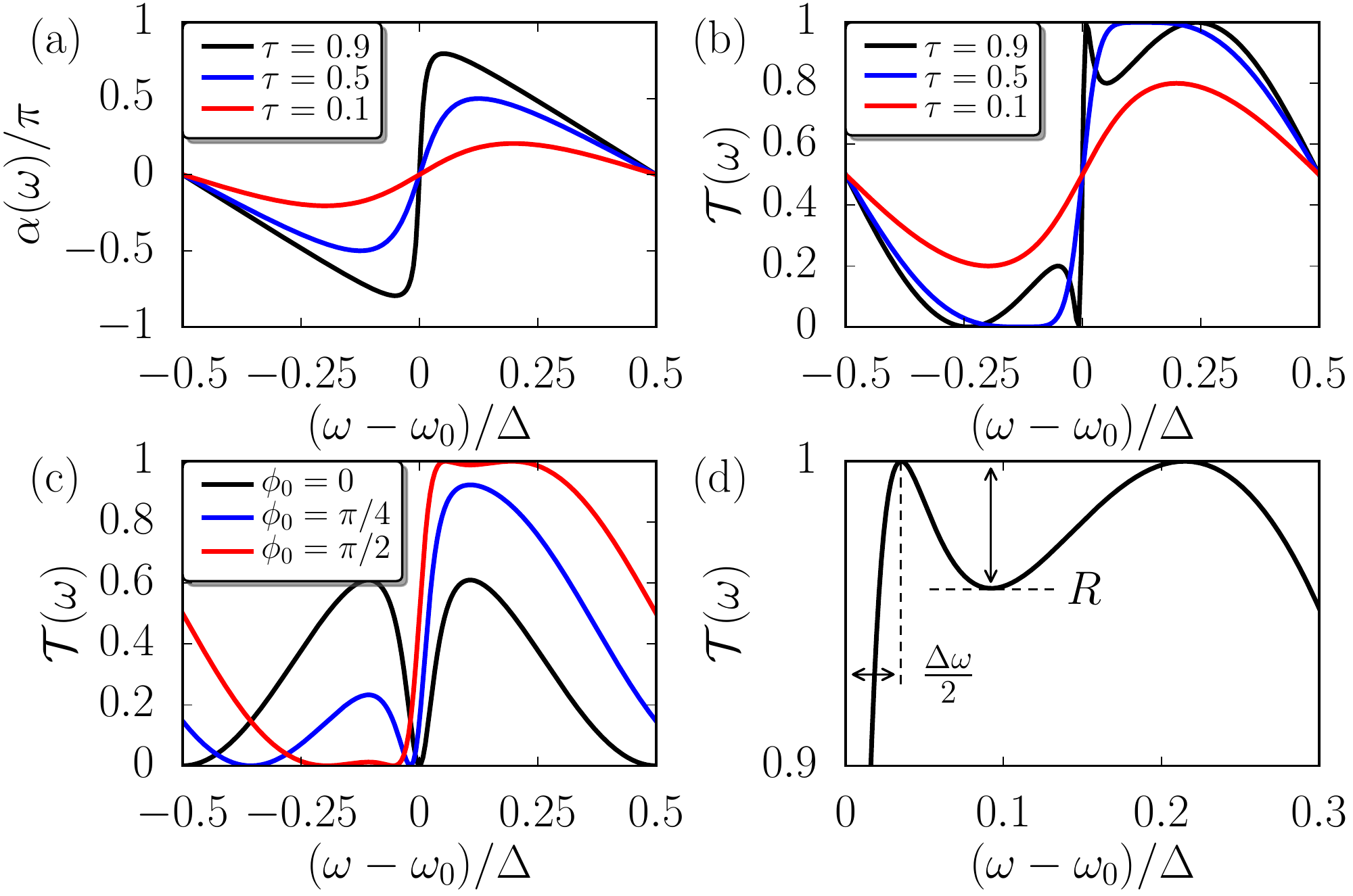}
	\caption{\label{fig:trans} (a) Phase $\alpha(\omega)$ for different transparencies $\tau$ of the capacitor-edge coupling. (b) Interferometer transmission probability ${\mathcal T}(\omega)$ for $\phi_0=\pi/2$ and same $\tau$ as in (a). (c) Transmission probability ${\mathcal T}(\omega)$ for $\tau=0.61$ and different $\phi_0$. (d) Close-up of ${\mathcal T}(\omega)$  defining magnitude of ripple $R$ and transition width $\Delta \omega$ for effective energy filter.}
\end{figure}

Importantly, due to the chiral nature of edge state transport there is no backscattering at the beam splitters or at the mesoscopic capacitor and the electrons pass through the interferometer only once. The total amplitude to transmit from the hot to the cold reservoir can hence be written, up to an overall phase factor, 
\begin{equation}
t(\omega)=\sqrt{\tau_A\tau_B}-\sqrt{(1-\tau_A)(1-\tau_B)}e^{i[\alpha(\omega)+\phi_0]}.
\label{amplitude}
\end{equation}
Here $\phi_0$ is an energy independent phase difference accounting for the enclosed Aharonov-Bohm flux and scattering phases of the beam splitters. The phase $\alpha(\omega)$, picked up when scattering forward at the capacitor and formally corresponding to $\alpha_1(\omega)-\alpha_2(\omega)-\pi$ in Eq.~\eqref{prob}, is obtained from~\cite{gabelli_violation_2006} as
\begin{equation}
	\alpha(\omega)=2\arctan\frac{\sqrt{\tau}\sin\left(2\pi\frac{\omega-\omega_0}{\Delta}\right)}{1-\sqrt{\tau}\cos\left(2\pi\frac{\omega-\omega_0}{\Delta}\right)},
\end{equation}
where $\omega_0$ is a gate-controllable resonance position. The phase $\alpha(\omega)$ over one period in energy, $\Delta$, resembling a smoothed sawtooth curve, is shown in Fig.~\ref{fig:trans} for three different $\tau$. Importantly, while the phase shift over an entire period is zero, the effective phase shift across the resonance energy $\omega_0$, conveniently defined as $\Delta \phi\equiv\max_{\omega}\{\alpha(\omega)\}-\min_{\omega}\{\alpha(\omega)\}$, is given by $\Delta \phi=4\arctan[\sqrt{\tau/(1-\tau)}]$. Hence, for $\tau=1/2$ we have $\Delta \phi=\pi$, in line with condition i) for optimal performance, albeit the shift (min to max) takes place over a finite energy $(\Delta/\pi)\arccos(\sqrt{\tau})= \Delta/4$.

To arrive at a transmission probability fulfilling condition ii) it is clear from Eqs.~\eqref{amplitude} and \eqref{prob} that we require semitransparent beam-splitters $\tau_A=\tau_B=1/2$. The total transmission probability ${\mathcal T}(\omega)=|t(\omega)|^2$ through the interferometer then becomes
\begin{equation}
{\mathcal T}(\omega)=\frac{\left[\sin \left(\frac{\phi_0}{2}\right)-\sqrt{\tau}\sin \left(\frac{\phi_0}{2}-2\pi\frac{\omega-\omega_0}{\Delta}\right)\right]^2}{1-2\sqrt{\tau}\cos \left(2\pi\frac{\omega-\omega_0}{\Delta}\right)+\tau}.
\end{equation}
The transmission probability is plotted in Fig.~\ref{fig:trans} for a set of different transparencies $\tau$ and phases $\phi_0$. Notably, for $\phi_0=0$ we have a symmetric transmission probability ${\mathcal T}(\omega-\omega_0)=\mathcal T(-[\omega-\omega_0])$ around the resonance, giving zero thermoelectric response for electron-hole symmetry $\omega_0=0$. In contrast, for $\phi_0=\pi/2$ the transmission is antisymmetric, ${\mathcal T}(\omega-\omega_0)=1-\mathcal T(-[\omega-\omega_0])$. In fact, for $\tau \geq 1/2$ and $\phi_0=\pi/2$ the transmission probability ${\mathcal T}(\omega)$ in a broad energy interval $\sim \Delta/2$ around the resonance, behaves as a low-pass energy filter. To quantify the deviations of ${\mathcal T}(\omega)$ from an ideal step function it is thus helpful to adopt a filter language~\cite{oppenheim_discrete-time_2009} and introduce (see lower right panel in Fig.~\ref{fig:trans} for definition), the ripple magnitude $R=1/2-\sqrt{\tau(1-\tau)}$ and the roll-off, or transition width $\Delta \omega=(\Delta/\pi)[\arcsin(1/\sqrt{2\tau})-\pi/4]$, unique functions of $\tau \geq 1/2$. Hence, increasing (reducing) $R$ leads to a smaller (larger) $\Delta \omega$. To find the optimal trade-off between $R$ and $\Delta \omega$, i.e., the optimal value of $\tau$, while at the same time identifying $\phi_0, \Delta$ and $\omega_0$ which optimize the interferometer thermoelectric performance, we turn to a quantitative analysis of the thermoelectric transport properties.

\paragraph{Scattering matrix theory.--}
Transport of charge and heat through a mesoscopic conductor can be described within the framework of scattering matrix theory~\cite{buttiker_absence_1988,butcher_thermal_1990}. For the two-terminal geometry in Fig.~\ref{fig:setup}, within linear response, we have $\vec I=\boldsymbol{\mathcal L}\vec F$. Here, $\vec I=\left(I_e,I_h\right)$ denotes the vector of charge and heat currents. The vector of thermodynamic forces is given by $\vec F=(F_V,F_T)$, where $F_V=eV/\kBT$ and $F_T=\kB\Delta T/(\kBT)^2$ with the bias voltage $V=V_C-V_H$ and the temperature bias $\Delta T=T_H-T_C$ applied between the two terminals, taking the cold terminal temperature equal to the background temperature, $T_C=T$. Finally, the Onsager matrix $\boldsymbol{\mathcal L}$ is given by
\begin{equation}
	\boldsymbol{\mathcal L}=\left(\begin{array}{cc}\mathcal L_{eV} & \mathcal L_{eT} \\ \mathcal L_{hV} & \mathcal L_{hT}\end{array}\right)=\frac{1}{h}\int d\omega\; \mathcal T(\omega)\xi(\omega)\left(\begin{array}{cc} e & e\omega \\ \omega & \omega^2\end{array}\right),
\end{equation}
where $\xi(\omega)=\left(2\cosh\frac{\omega}{2\kBT}\right)^{-2}$. The diagonal Onsager coefficients $\mathcal L_{eV}$ and $\mathcal L_{hT}$ are related to the electrical and thermal conductance, respectively, while the off-diagonal ones, $\mathcal L_{eT}$ and $\mathcal L_{hV}$, are linked to the Seebeck and Peltier coefficients.

Applying a temperature bias $\Delta T>0$ to the setup, we can drive a charge current through the system, against the externally applied bias voltage $V>0$. Optimizing the bias voltage for a given temperature bias, we obtain the maximal output power~\cite{benenti_thermodynamic_2011} 
\begin{equation}\label{eq:Pmax}
	P_\text{max}=\frac{\kBT}{4e}\frac{(\mathcal L_{eT})^2}{\mathcal L_{eV}} (F_T)^2.
\end{equation}
The associated efficiency at maximum power, defined as the ratio between output power $P_\text{max}$ and input heat $I_h$, is given by~\cite{benenti_thermodynamic_2011}
\begin{equation}\label{eq:etamaxP}
	\eta_\text{maxP}=\frac{\eta_C}{2e}\frac{(\mathcal L_{eT})^2}{2\mathcal L_{eV}\mathcal L_{hT}-\mathcal L_{eT}\mathcal L_{hV}},
\end{equation}
where $\eta_C=\Delta T/T$ denotes the Carnot efficiency in linear response. 

\paragraph{Close-to-optimal performance.--}
To benchmark the performance of our interferometer we first recall~\cite{whitney_most_2014,whitney_finding_2015} the theoretical, maximum single mode power for a heat engine, obtained for a step-function transmission probability $\mathcal T(\omega)=\theta(\omega-\omega_0)$. The output power is bounded by $P_\text{max}=0.32(\kB\Delta T)^2/h$,  obtained for $\omega_0=1.16k_{\mathrm{B}}T$. The corresponding efficiency is given by $\eta_\text{maxP}=0.35\eta_C$. We stress that for a transmission probability different from a step function, giving a smaller maximum power, the corresponding efficiency can reach a larger fraction of $\eta_C$. The efficiency at maximum power is ultimately bounded by the Curzon-Ahlborn limit of $\eta_C/2$, obtained for a $\delta$-function $\mathcal T(\omega)$ in the limit $\tau\to1$, $\phi_0=\pi$ for which the maximum power approaches zero.
\begin{figure}
	\includegraphics[width=\columnwidth]{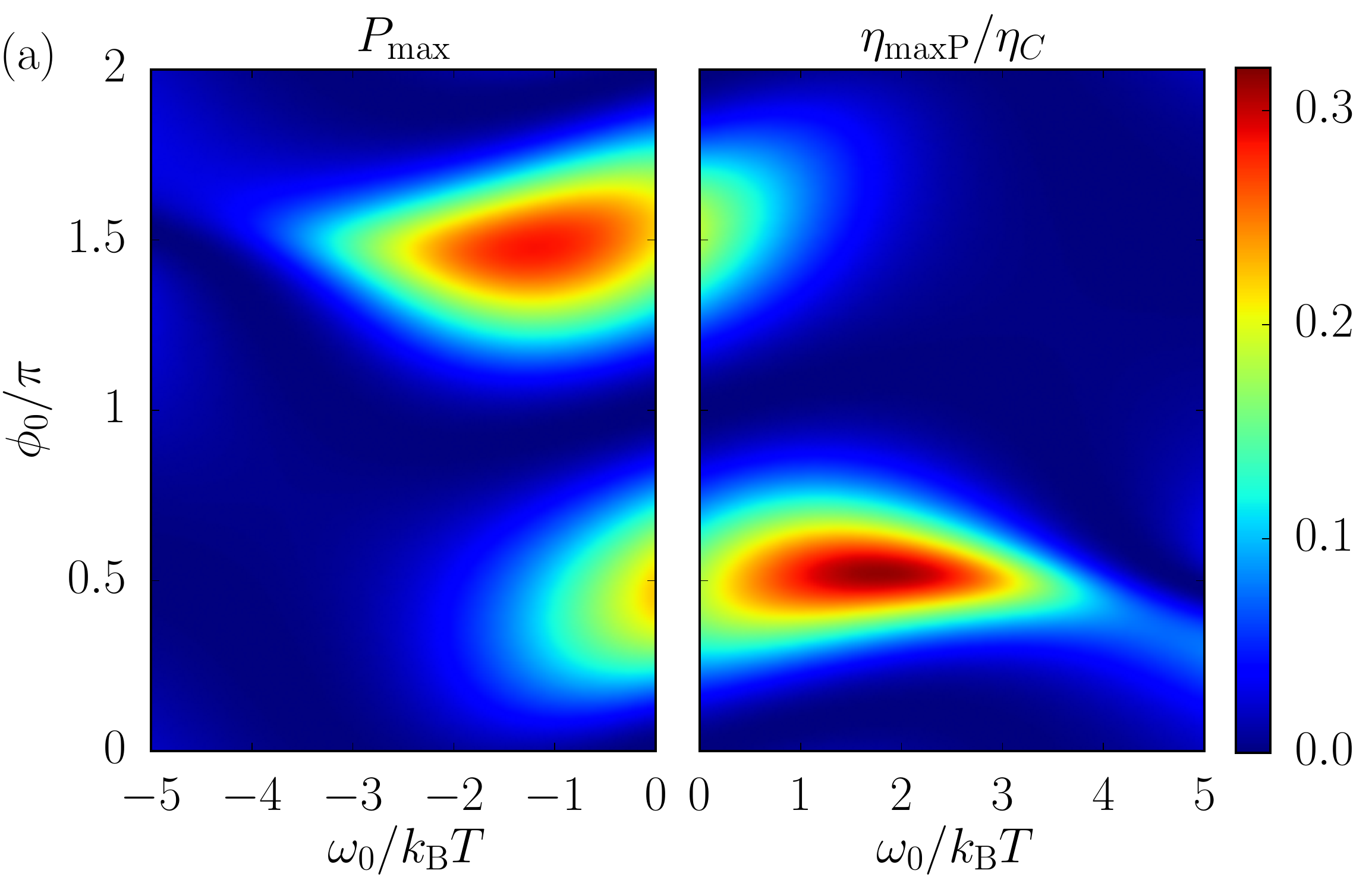}
	\includegraphics[width=.49\columnwidth]{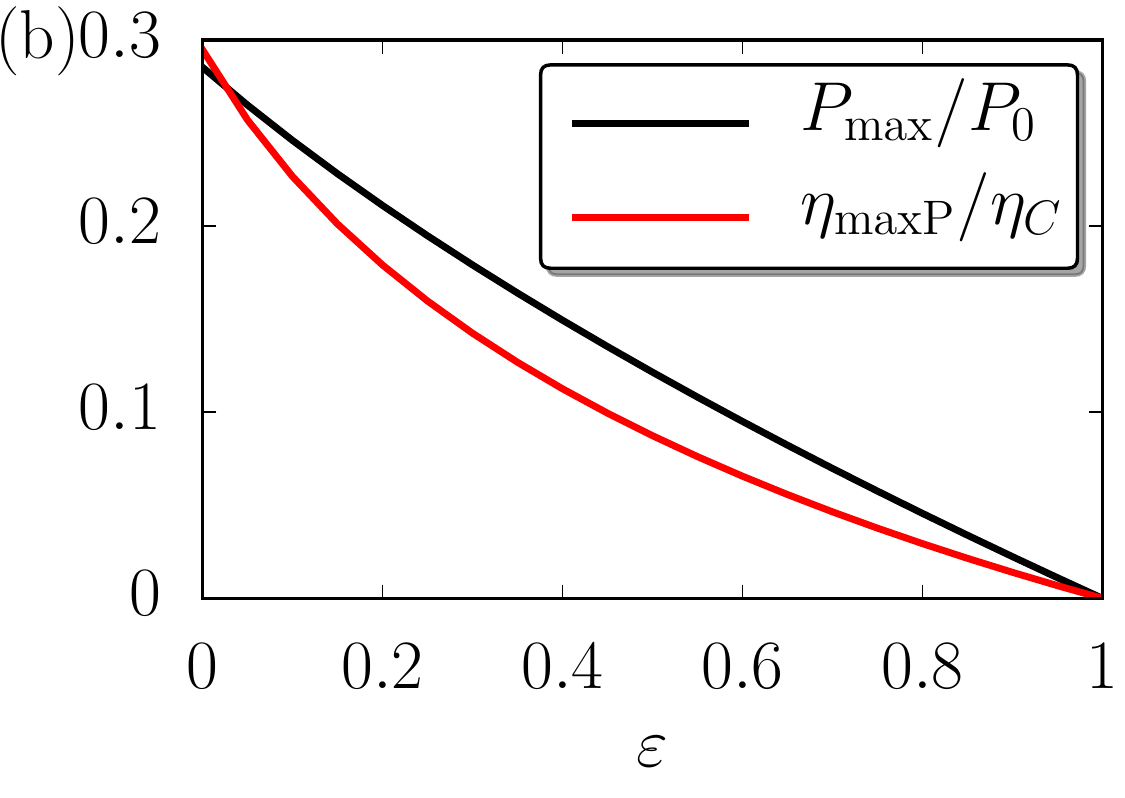}
	\includegraphics[width=.49\columnwidth]{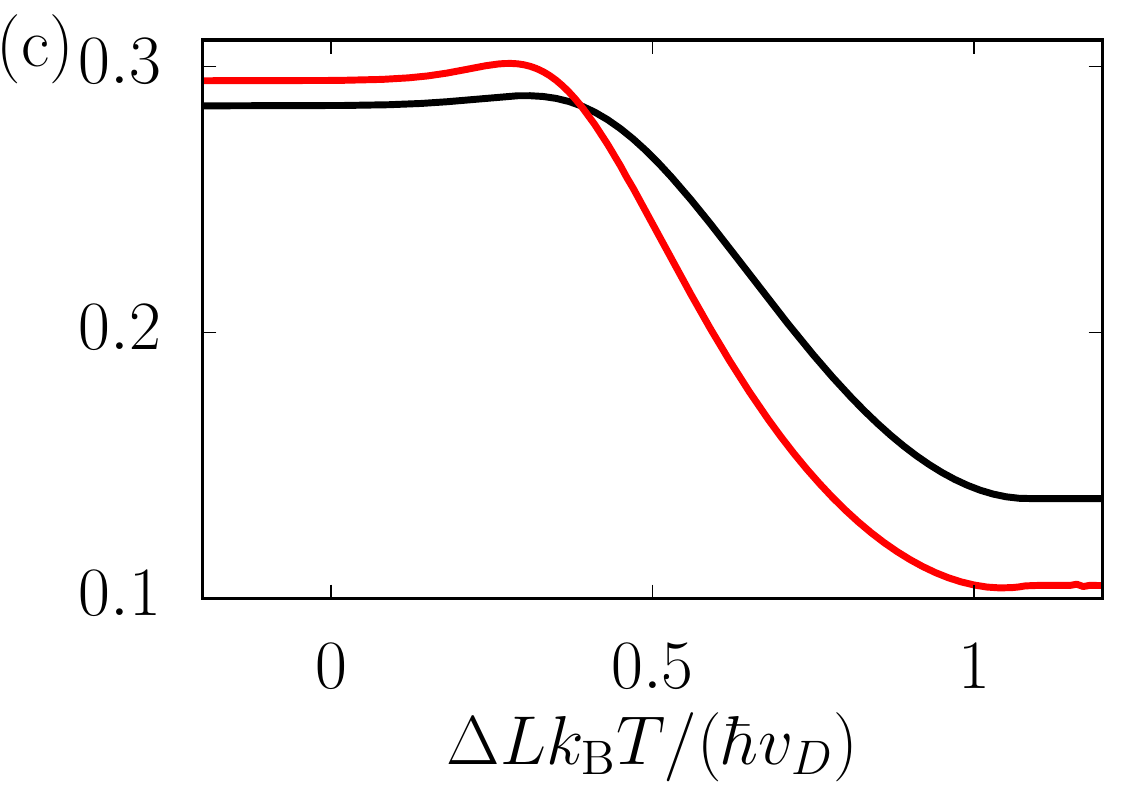}
	\caption{\label{fig:PowerEfficiency} (a) Maximal power in units of $P_0=(\kB\Delta T)^2/h$ (left) and efficiency at maximum power in units of $\eta_C$ (right) as functions of $\phi_0$ and $\omega_0$. Parameters are $\Delta=24.2\kBT$ and $\tau=0.61$. (b) Maximum power in units of $P_0$ and efficiency at maximum power in units of $\eta_C$, as a functions of dephasing strength $0 \leq \epsilon \leq 1$ and (c) effective arm length difference $\Delta L k_{\mathrm{B}}T/(\hbar v_{\mathrm{D}})$.}
\end{figure}

To obtain the maximum power of our interferometer we perform a numerical optimization of $P_{\mathrm{max}}$ in Eq.~\eqref{eq:Pmax} over $0 \leq \tau \leq 1, 0\leq \phi_0 \leq \pi, \omega_0$ and $\Delta$ for a given background temperature $T$. In this way, we find that the maximum power becomes largest for $\tau=0.61$ and $\phi_0=0.52\pi$, together with $\omega_0=1.17\kBT$ and $\Delta=24.2\kBT$. For these particular parameters, we have $P_\text{max}=0.285(\kB\Delta T)^2/h$, which reaches $90\%$ of the optimal value. The corresponding efficiency at maximum power is given by $\eta_\text{maxP}=0.29\eta_C$, which is $83\%$ of its upper bound. We thus have that our heat engine outperforms the previously suggested Mach-Zehnder heat engine~\cite{hofer_quantum_2015}, with a linear-in-energy dependence of the phase $\alpha(\omega)\propto \omega$ due to arm-length asymmetry,  by a factor of 2 in terms of power output and by a factor of 3 in terms of the efficiency at maximum power.

We point out that the obtained value of $\phi_0$ is close to $\pi/2$, identified as ideal from the qualitative analysis above. The optimal value of $\tau$ corresponds to a ripple magnitude $R\approx0.01$, and a roll-off $\Delta\omega\approx0.11\Delta$, indicating that a smooth pass and stop band is more important for an optimal device performance than the transition width.  Moreover, the maximum power and the efficiency at maximum power for the obtained $\tau=0.61$ and $\Delta=24.2\kBT$ are plotted as functions of $\phi_0$ and $\omega_0/\kBT$ in Fig.~\ref{fig:PowerEfficiency}, displaying broad parameter regions with large $P_{\mathrm{max}}$ and $\eta_\text{maxP}$ around $\omega_0=1.17\kBT$ and $\phi_0=0.52\pi$. 

\paragraph{Dephasing and path-length difference.--}
A key question is how robust the obtained results are to possible deviations from optimal conditions, most importantly due to dephasing effects and unequal path lengths. Dephasing, the loss of phase coherence for electrons traversing the interferometers, has been investigated in depth for Mach-Zehnder interferometers \cite{neder_controlled_2007,neder_entanglement_2007,roulleau_direct_2008,roulleau_noise_2008,roulleau_tuning_2009,huynh_quantum_2012,helzel_counting_2015}. Adapting here a  qualitative model~\cite{pilgram_full-counting_2006,forster_voltage_2007}, treating dephasing as phase averaging or equivalently, via coupling of the interferometer to a dephasing probe, we can write the total phase as $\phi_0=\bar \phi_0+\delta \phi$, a sum of a constant, controllable part $\bar \phi_0$ and a randomly fluctuating part $\delta \phi$. The $2\pi$-periodic distribution of the fluctuating part is $\rho(\delta \phi)=(1/2\pi)\epsilon/[2-\epsilon-2\sqrt{1-\epsilon}\cos(\delta \phi)]$ where $0 \leq \epsilon \leq 1$ characterizes the strength of the dephasing. We note that for a Mach-Zehnder interferometer without the capacitor the visibility of the phase oscillations~\cite{chung_visibility_2005} is $\sqrt{1-\epsilon}$, i.e., the observed $90\%$ visibility~\cite{neder_interference_2007} corresponds to $\epsilon=0.2$. 

The transmission probability in the presence of dephasing, ${\mathcal T}_{\mathrm{deph}}(\omega)=\int d\delta \phi \rho(\delta \phi){\mathcal T}(\omega,\phi_0)$, is given by 
\begin{equation}
{\mathcal T}_{\mathrm{deph}}(\omega)=\sqrt{1-\epsilon}{\mathcal T}(\omega,\bar \phi_0)+(1-\sqrt{1-\epsilon})/2,
\end{equation}
i.e., dephasing effectively suppresses the energy dependence of the transmission probability. Performing a numerical optimization of the maximum power and efficiency at maximum power in the presence of dephasing we find, see Fig.~\ref{fig:PowerEfficiency}, that $P_{\mathrm{max}}$ and $\eta_\text{maxP}$ decrease slowly, close to linearly with increasing dephasing strength $\epsilon$. Hence, the predicted thermoelectric effects are robust to moderate dephasing. Moreover (not shown), the parameters giving optimal performance are largely unaffected by the dephasing.
Note that for relevant parameters $\tau>0.1$ and a capacitor's circumference much shorter than $L_1+L_2$, the additional dephasing inside the capacitor is small compared to the overall interferometer dephasing and hence neglected~\cite{parmentier_current_2012}.
We stress that for complete dephasing, $\epsilon=1$, we have ${\mathcal T}(\omega,\bar \phi_0)=1/2$, independent on energy $\omega$ and $\bar \phi_0$, and hence zero thermoelectric response as expected from a heat engine operating only due to quantum interference.

Unequal interferometer arm lengths, $\Delta L \equiv L_1-(L_2+L) \neq 0$, can formally be incorporated in the transmission probability $\mathcal T(\omega)$ in Eq. \eqref{prob} by changing $\phi_0 \rightarrow \phi_0 +\omega \Delta L/(\hbar v_\text{D})$. The result of a numerical optimization of $P_{\mathrm{max}}$ and $\eta_\text{maxP}$ (for zero dephasing, $\epsilon=0$) is shown in Fig.~\ref{fig:PowerEfficiency}. Interestingly, by tuning $\Delta L \sim 0.35 \hbar v_\text{D}/(\kBT)$ it is possible to marginally increase the thermoelectric performance. For $\Delta L \sim -0.1 \hbar v_\text{D}/(\kBT)$ the maximizing level spacing diverges, $\Delta/\kBT \rightarrow \infty$, while for $\Delta L \sim \hbar v_\text{D}/(\kBT)$ the optimizing transmission reaches $\tau=1$ (or $\tau=0$) and we recover the Mach-Zehnder result of Ref.~\cite{hofer_mach-zehnder_2014}.  

\paragraph{Summary.--}
We have proposed a heat engine operating uniquely due to quantum-mechanical interference and derived fundamental limits on its maximal output power and efficiency at maximum power.
We have suggested an experimentally feasible realization based on an electronic Mach-Zehnder interferometer with a side-coupled mesoscopic capacitor. The setup saturates the general power and efficiency bounds at 90\% and 83\%, respectively. Driven by a realistic temperature bias of $\unit[50]{mK}$, it can deliver a current of $\unit[0.1]{nA}$ together with an output power of $\unit[0.2]{fW}$. Our findings contribute significantly to the fundamental investigations of quantum coherence in the performance of thermodynamic devices~\cite{streltsov_quantum_2016} and provide a means for harnessing quantum interference for experimentally optimizing nanoscale heat engines.

\acknowledgments
\paragraph{Acknowledgments.--} We acknowledge discussions with F. Brange, G. Fève, P. P. Hofer, M. Leijnse, R. Sánchez, J. Splettst\"osser and A. Wacker. We acknowledge financial support from the Ministry of Innovation NRW, the COST Action MP1209, the Swedish VR and the EU ITN PhD4Energy, Grant No. 608153.


\begin{thebibliography}{72}%
\makeatletter
\providecommand \@ifxundefined [1]{%
 \@ifx{#1\undefined}
}%
\providecommand \@ifnum [1]{%
 \ifnum #1\expandafter \@firstoftwo
 \else \expandafter \@secondoftwo
 \fi
}%
\providecommand \@ifx [1]{%
 \ifx #1\expandafter \@firstoftwo
 \else \expandafter \@secondoftwo
 \fi
}%
\providecommand \natexlab [1]{#1}%
\providecommand \enquote  [1]{``#1''}%
\providecommand \bibnamefont  [1]{#1}%
\providecommand \bibfnamefont [1]{#1}%
\providecommand \citenamefont [1]{#1}%
\providecommand \href@noop [0]{\@secondoftwo}%
\providecommand \href [0]{\begingroup \@sanitize@url \@href}%
\providecommand \@href[1]{\@@startlink{#1}\@@href}%
\providecommand \@@href[1]{\endgroup#1\@@endlink}%
\providecommand \@sanitize@url [0]{\catcode `\\12\catcode `\$12\catcode
  `\&12\catcode `\#12\catcode `\^12\catcode `\_12\catcode `\%12\relax}%
\providecommand \@@startlink[1]{}%
\providecommand \@@endlink[0]{}%
\providecommand \url  [0]{\begingroup\@sanitize@url \@url }%
\providecommand \@url [1]{\endgroup\@href {#1}{\urlprefix }}%
\providecommand \urlprefix  [0]{URL }%
\providecommand \Eprint [0]{\href }%
\providecommand \doibase [0]{http://dx.doi.org/}%
\providecommand \selectlanguage [0]{\@gobble}%
\providecommand \bibinfo  [0]{\@secondoftwo}%
\providecommand \bibfield  [0]{\@secondoftwo}%
\providecommand \translation [1]{[#1]}%
\providecommand \BibitemOpen [0]{}%
\providecommand \bibitemStop [0]{}%
\providecommand \bibitemNoStop [0]{.\EOS\space}%
\providecommand \EOS [0]{\spacefactor3000\relax}%
\providecommand \BibitemShut  [1]{\csname bibitem#1\endcsname}%
\let\auto@bib@innerbib\@empty
\bibitem [{\citenamefont {Shakouri}(2011)}]{shakouri_recent_2011}%
  \BibitemOpen
  \bibfield  {author} {\bibinfo {author} {\bibfnamefont {A.}~\bibnamefont
  {Shakouri}},\ }\href {\doibase 10.1146/annurev-matsci-062910-100445}
  {\bibfield  {journal} {\bibinfo  {journal} {Annu. Rev. Mater. Res.}\ }\textbf
  {\bibinfo {volume} {41}},\ \bibinfo {pages} {399} (\bibinfo {year}
  {2011})}\BibitemShut {NoStop}%
\bibitem [{\citenamefont {Streda}(1989)}]{streda_quantised_1989}%
  \BibitemOpen
  \bibfield  {author} {\bibinfo {author} {\bibfnamefont {P.}~\bibnamefont
  {Streda}},\ }\href {\doibase 10.1088/0953-8984/1/5/021} {\bibfield  {journal}
  {\bibinfo  {journal} {J. Phys.: Condens. Matter}\ }\textbf {\bibinfo {volume}
  {1}},\ \bibinfo {pages} {1025} (\bibinfo {year} {1989})}\BibitemShut
  {NoStop}%
\bibitem [{\citenamefont {Beenakker}\ and\ \citenamefont
  {Staring}(1992)}]{beenakker_theory_1992}%
  \BibitemOpen
  \bibfield  {author} {\bibinfo {author} {\bibfnamefont {C.~W.~J.}\
  \bibnamefont {Beenakker}}\ and\ \bibinfo {author} {\bibfnamefont {A.~A.~M.}\
  \bibnamefont {Staring}},\ }\href {\doibase 10.1103/PhysRevB.46.9667}
  {\bibfield  {journal} {\bibinfo  {journal} {Phys. Rev. B}\ }\textbf {\bibinfo
  {volume} {46}},\ \bibinfo {pages} {9667} (\bibinfo {year}
  {1992})}\BibitemShut {NoStop}%
\bibitem [{\citenamefont {Hicks}\ and\ \citenamefont
  {Dresselhaus}(1993{\natexlab{a}})}]{hicks_effect_1993}%
  \BibitemOpen
  \bibfield  {author} {\bibinfo {author} {\bibfnamefont {L.~D.}\ \bibnamefont
  {Hicks}}\ and\ \bibinfo {author} {\bibfnamefont {M.~S.}\ \bibnamefont
  {Dresselhaus}},\ }\href {\doibase 10.1103/PhysRevB.47.12727} {\bibfield
  {journal} {\bibinfo  {journal} {Phys. Rev. B}\ }\textbf {\bibinfo {volume}
  {47}},\ \bibinfo {pages} {12727} (\bibinfo {year}
  {1993}{\natexlab{a}})}\BibitemShut {NoStop}%
\bibitem [{\citenamefont {Hicks}\ and\ \citenamefont
  {Dresselhaus}(1993{\natexlab{b}})}]{hicks_thermoelectric_1993}%
  \BibitemOpen
  \bibfield  {author} {\bibinfo {author} {\bibfnamefont {L.~D.}\ \bibnamefont
  {Hicks}}\ and\ \bibinfo {author} {\bibfnamefont {M.~S.}\ \bibnamefont
  {Dresselhaus}},\ }\href {\doibase 10.1103/PhysRevB.47.16631} {\bibfield
  {journal} {\bibinfo  {journal} {Phys. Rev. B}\ }\textbf {\bibinfo {volume}
  {47}},\ \bibinfo {pages} {16631} (\bibinfo {year}
  {1993}{\natexlab{b}})}\BibitemShut {NoStop}%
\bibitem [{\citenamefont {Nakpathomkun}\ \emph {et~al.}(2010)\citenamefont
  {Nakpathomkun}, \citenamefont {Xu},\ and\ \citenamefont
  {Linke}}]{nakpathomkun_thermoelectric_2010}%
  \BibitemOpen
  \bibfield  {author} {\bibinfo {author} {\bibfnamefont {N.}~\bibnamefont
  {Nakpathomkun}}, \bibinfo {author} {\bibfnamefont {H.~Q.}\ \bibnamefont
  {Xu}}, \ and\ \bibinfo {author} {\bibfnamefont {H.}~\bibnamefont {Linke}},\
  }\href {\doibase 10.1103/PhysRevB.82.235428} {\bibfield  {journal} {\bibinfo
  {journal} {Phys. Rev. B}\ }\textbf {\bibinfo {volume} {82}},\ \bibinfo
  {pages} {235428} (\bibinfo {year} {2010})}\BibitemShut {NoStop}%
\bibitem [{\citenamefont {Karlström}\ \emph {et~al.}(2011)\citenamefont
  {Karlström}, \citenamefont {Linke}, \citenamefont {Karlström},\ and\
  \citenamefont {Wacker}}]{karlstrom_increasing_2011}%
  \BibitemOpen
  \bibfield  {author} {\bibinfo {author} {\bibfnamefont {O.}~\bibnamefont
  {Karlström}}, \bibinfo {author} {\bibfnamefont {H.}~\bibnamefont {Linke}},
  \bibinfo {author} {\bibfnamefont {G.}~\bibnamefont {Karlström}}, \ and\
  \bibinfo {author} {\bibfnamefont {A.}~\bibnamefont {Wacker}},\ }\href
  {\doibase 10.1103/PhysRevB.84.113415} {\bibfield  {journal} {\bibinfo
  {journal} {Phys. Rev. B}\ }\textbf {\bibinfo {volume} {84}},\ \bibinfo
  {pages} {113415} (\bibinfo {year} {2011})}\BibitemShut {NoStop}%
\bibitem [{\citenamefont {Molenkamp}\ \emph {et~al.}(1990)\citenamefont
  {Molenkamp}, \citenamefont {van Houten}, \citenamefont {Beenakker},
  \citenamefont {Eppenga},\ and\ \citenamefont
  {Foxon}}]{molenkamp_quantum_1990}%
  \BibitemOpen
  \bibfield  {author} {\bibinfo {author} {\bibfnamefont {L.~W.}\ \bibnamefont
  {Molenkamp}}, \bibinfo {author} {\bibfnamefont {H.}~\bibnamefont {van
  Houten}}, \bibinfo {author} {\bibfnamefont {C.~W.~J.}\ \bibnamefont
  {Beenakker}}, \bibinfo {author} {\bibfnamefont {R.}~\bibnamefont {Eppenga}},
  \ and\ \bibinfo {author} {\bibfnamefont {C.~T.}\ \bibnamefont {Foxon}},\
  }\href {\doibase 10.1103/PhysRevLett.65.1052} {\bibfield  {journal} {\bibinfo
   {journal} {Phys. Rev. Lett.}\ }\textbf {\bibinfo {volume} {65}},\ \bibinfo
  {pages} {1052} (\bibinfo {year} {1990})}\BibitemShut {NoStop}%
\bibitem [{\citenamefont {Staring}\ \emph {et~al.}(1993)\citenamefont
  {Staring}, \citenamefont {Molenkamp}, \citenamefont {Alphenaar},
  \citenamefont {van Houten}, \citenamefont {Buyk}, \citenamefont {Mabesoone},
  \citenamefont {Beenakker},\ and\ \citenamefont
  {Foxon}}]{staring_coulomb-blockade_1993}%
  \BibitemOpen
  \bibfield  {author} {\bibinfo {author} {\bibfnamefont {A.~A.~M.}\
  \bibnamefont {Staring}}, \bibinfo {author} {\bibfnamefont {L.~W.}\
  \bibnamefont {Molenkamp}}, \bibinfo {author} {\bibfnamefont {B.~W.}\
  \bibnamefont {Alphenaar}}, \bibinfo {author} {\bibfnamefont {H.}~\bibnamefont
  {van Houten}}, \bibinfo {author} {\bibfnamefont {O.~J.~A.}\ \bibnamefont
  {Buyk}}, \bibinfo {author} {\bibfnamefont {M.~A.~A.}\ \bibnamefont
  {Mabesoone}}, \bibinfo {author} {\bibfnamefont {C.~W.~J.}\ \bibnamefont
  {Beenakker}}, \ and\ \bibinfo {author} {\bibfnamefont {C.~T.}\ \bibnamefont
  {Foxon}},\ }\href {\doibase 10.1209/0295-5075/22/1/011} {\bibfield  {journal}
  {\bibinfo  {journal} {Europhysics Letters (EPL)}\ }\textbf {\bibinfo {volume}
  {22}},\ \bibinfo {pages} {57} (\bibinfo {year} {1993})}\BibitemShut {NoStop}%
\bibitem [{\citenamefont {Scheibner}\ \emph {et~al.}(2005)\citenamefont
  {Scheibner}, \citenamefont {Buhmann}, \citenamefont {Reuter}, \citenamefont
  {Kiselev},\ and\ \citenamefont {Molenkamp}}]{scheibner_thermopower_2005}%
  \BibitemOpen
  \bibfield  {author} {\bibinfo {author} {\bibfnamefont {R.}~\bibnamefont
  {Scheibner}}, \bibinfo {author} {\bibfnamefont {H.}~\bibnamefont {Buhmann}},
  \bibinfo {author} {\bibfnamefont {D.}~\bibnamefont {Reuter}}, \bibinfo
  {author} {\bibfnamefont {M.~N.}\ \bibnamefont {Kiselev}}, \ and\ \bibinfo
  {author} {\bibfnamefont {L.~W.}\ \bibnamefont {Molenkamp}},\ }\href {\doibase
  10.1103/PhysRevLett.95.176602} {\bibfield  {journal} {\bibinfo  {journal}
  {Phys. Rev. Lett.}\ }\textbf {\bibinfo {volume} {95}},\ \bibinfo {pages}
  {176602} (\bibinfo {year} {2005})}\BibitemShut {NoStop}%
\bibitem [{\citenamefont {Svensson}\ \emph {et~al.}(2012)\citenamefont
  {Svensson}, \citenamefont {Persson}, \citenamefont {Hoffmann}, \citenamefont
  {Nakpathomkun}, \citenamefont {Nilsson}, \citenamefont {Xu}, \citenamefont
  {Samuelson},\ and\ \citenamefont {Linke}}]{svensson_lineshape_2012}%
  \BibitemOpen
  \bibfield  {author} {\bibinfo {author} {\bibfnamefont {S.~F.}\ \bibnamefont
  {Svensson}}, \bibinfo {author} {\bibfnamefont {A.~I.}\ \bibnamefont
  {Persson}}, \bibinfo {author} {\bibfnamefont {E.~A.}\ \bibnamefont
  {Hoffmann}}, \bibinfo {author} {\bibfnamefont {N.}~\bibnamefont
  {Nakpathomkun}}, \bibinfo {author} {\bibfnamefont {H.~A.}\ \bibnamefont
  {Nilsson}}, \bibinfo {author} {\bibfnamefont {H.~Q.}\ \bibnamefont {Xu}},
  \bibinfo {author} {\bibfnamefont {L.}~\bibnamefont {Samuelson}}, \ and\
  \bibinfo {author} {\bibfnamefont {H.}~\bibnamefont {Linke}},\ }\href
  {\doibase 10.1088/1367-2630/14/3/033041} {\bibfield  {journal} {\bibinfo
  {journal} {New J. Phys.}\ }\textbf {\bibinfo {volume} {14}},\ \bibinfo
  {pages} {033041} (\bibinfo {year} {2012})}\BibitemShut {NoStop}%
\bibitem [{\citenamefont {Svensson}\ \emph {et~al.}(2013)\citenamefont
  {Svensson}, \citenamefont {Hoffmann}, \citenamefont {Nakpathomkun},
  \citenamefont {Wu}, \citenamefont {Xu}, \citenamefont {Nilsson},
  \citenamefont {Sánchez}, \citenamefont {Kashcheyevs},\ and\ \citenamefont
  {Linke}}]{svensson_nonlinear_2013}%
  \BibitemOpen
  \bibfield  {author} {\bibinfo {author} {\bibfnamefont {S.~F.}\ \bibnamefont
  {Svensson}}, \bibinfo {author} {\bibfnamefont {E.~A.}\ \bibnamefont
  {Hoffmann}}, \bibinfo {author} {\bibfnamefont {N.}~\bibnamefont
  {Nakpathomkun}}, \bibinfo {author} {\bibfnamefont {P.~M.}\ \bibnamefont
  {Wu}}, \bibinfo {author} {\bibfnamefont {H.~Q.}\ \bibnamefont {Xu}}, \bibinfo
  {author} {\bibfnamefont {H.~A.}\ \bibnamefont {Nilsson}}, \bibinfo {author}
  {\bibfnamefont {D.}~\bibnamefont {Sánchez}}, \bibinfo {author}
  {\bibfnamefont {V.}~\bibnamefont {Kashcheyevs}}, \ and\ \bibinfo {author}
  {\bibfnamefont {H.}~\bibnamefont {Linke}},\ }\href {\doibase
  10.1088/1367-2630/15/10/105011} {\bibfield  {journal} {\bibinfo  {journal}
  {New J. Phys.}\ }\textbf {\bibinfo {volume} {15}},\ \bibinfo {pages} {105011}
  (\bibinfo {year} {2013})}\BibitemShut {NoStop}%
\bibitem [{\citenamefont {Mahan}\ and\ \citenamefont
  {Sofo}(1996)}]{mahan_best_1996}%
  \BibitemOpen
  \bibfield  {author} {\bibinfo {author} {\bibfnamefont {G.~D.}\ \bibnamefont
  {Mahan}}\ and\ \bibinfo {author} {\bibfnamefont {J.~O.}\ \bibnamefont
  {Sofo}},\ }\href {http://www.pnas.org/content/93/15/7436} {\bibfield
  {journal} {\bibinfo  {journal} {Proc. Natl. Acad. Sci. USA}\ }\textbf
  {\bibinfo {volume} {93}},\ \bibinfo {pages} {7436} (\bibinfo {year}
  {1996})}\BibitemShut {NoStop}%
\bibitem [{\citenamefont {Entin-Wohlman}\ \emph {et~al.}(2010)\citenamefont
  {Entin-Wohlman}, \citenamefont {Imry},\ and\ \citenamefont
  {Aharony}}]{entin-wohlman_three-terminal_2010}%
  \BibitemOpen
  \bibfield  {author} {\bibinfo {author} {\bibfnamefont {O.}~\bibnamefont
  {Entin-Wohlman}}, \bibinfo {author} {\bibfnamefont {Y.}~\bibnamefont {Imry}},
  \ and\ \bibinfo {author} {\bibfnamefont {A.}~\bibnamefont {Aharony}},\ }\href
  {\doibase 10.1103/PhysRevB.82.115314} {\bibfield  {journal} {\bibinfo
  {journal} {Phys. Rev. B}\ }\textbf {\bibinfo {volume} {82}},\ \bibinfo
  {pages} {115314} (\bibinfo {year} {2010})}\BibitemShut {NoStop}%
\bibitem [{\citenamefont {Sánchez}\ and\ \citenamefont
  {Büttiker}(2011)}]{sanchez_optimal_2011}%
  \BibitemOpen
  \bibfield  {author} {\bibinfo {author} {\bibfnamefont {R.}~\bibnamefont
  {Sánchez}}\ and\ \bibinfo {author} {\bibfnamefont {M.}~\bibnamefont
  {Büttiker}},\ }\href {\doibase 10.1103/PhysRevB.83.085428} {\bibfield
  {journal} {\bibinfo  {journal} {Phys. Rev. B}\ }\textbf {\bibinfo {volume}
  {83}},\ \bibinfo {pages} {085428} (\bibinfo {year} {2011})}\BibitemShut
  {NoStop}%
\bibitem [{\citenamefont {Sothmann}\ \emph {et~al.}(2012)\citenamefont
  {Sothmann}, \citenamefont {Sánchez}, \citenamefont {Jordan},\ and\
  \citenamefont {Büttiker}}]{sothmann_rectification_2012}%
  \BibitemOpen
  \bibfield  {author} {\bibinfo {author} {\bibfnamefont {B.}~\bibnamefont
  {Sothmann}}, \bibinfo {author} {\bibfnamefont {R.}~\bibnamefont {Sánchez}},
  \bibinfo {author} {\bibfnamefont {A.~N.}\ \bibnamefont {Jordan}}, \ and\
  \bibinfo {author} {\bibfnamefont {M.}~\bibnamefont {Büttiker}},\ }\href
  {\doibase 10.1103/PhysRevB.85.205301} {\bibfield  {journal} {\bibinfo
  {journal} {Phys. Rev. B}\ }\textbf {\bibinfo {volume} {85}},\ \bibinfo
  {pages} {205301} (\bibinfo {year} {2012})}\BibitemShut {NoStop}%
\bibitem [{\citenamefont {Sothmann}\ and\ \citenamefont
  {Büttiker}(2012)}]{sothmann_magnon-driven_2012}%
  \BibitemOpen
  \bibfield  {author} {\bibinfo {author} {\bibfnamefont {B.}~\bibnamefont
  {Sothmann}}\ and\ \bibinfo {author} {\bibfnamefont {M.}~\bibnamefont
  {Büttiker}},\ }\href {\doibase 10.1209/0295-5075/99/27001} {\bibfield
  {journal} {\bibinfo  {journal} {Europhys. Lett.}\ }\textbf {\bibinfo {volume}
  {99}},\ \bibinfo {pages} {27001} (\bibinfo {year} {2012})}\BibitemShut
  {NoStop}%
\bibitem [{\citenamefont {Jiang}\ \emph {et~al.}(2013)\citenamefont {Jiang},
  \citenamefont {Entin-Wohlman},\ and\ \citenamefont
  {Imry}}]{jiang_three-terminal_2013}%
  \BibitemOpen
  \bibfield  {author} {\bibinfo {author} {\bibfnamefont {J.-H.}\ \bibnamefont
  {Jiang}}, \bibinfo {author} {\bibfnamefont {O.}~\bibnamefont
  {Entin-Wohlman}}, \ and\ \bibinfo {author} {\bibfnamefont {Y.}~\bibnamefont
  {Imry}},\ }\href {\doibase 10.1088/1367-2630/15/7/075021} {\bibfield
  {journal} {\bibinfo  {journal} {New J. Phys.}\ }\textbf {\bibinfo {volume}
  {15}},\ \bibinfo {pages} {075021} (\bibinfo {year} {2013})}\BibitemShut
  {NoStop}%
\bibitem [{\citenamefont {Machon}\ \emph {et~al.}(2013)\citenamefont {Machon},
  \citenamefont {Eschrig},\ and\ \citenamefont
  {Belzig}}]{machon_nonlocal_2013}%
  \BibitemOpen
  \bibfield  {author} {\bibinfo {author} {\bibfnamefont {P.}~\bibnamefont
  {Machon}}, \bibinfo {author} {\bibfnamefont {M.}~\bibnamefont {Eschrig}}, \
  and\ \bibinfo {author} {\bibfnamefont {W.}~\bibnamefont {Belzig}},\ }\href
  {\doibase 10.1103/PhysRevLett.110.047002} {\bibfield  {journal} {\bibinfo
  {journal} {Phys. Rev. Lett.}\ }\textbf {\bibinfo {volume} {110}},\ \bibinfo
  {pages} {047002} (\bibinfo {year} {2013})}\BibitemShut {NoStop}%
\bibitem [{\citenamefont {Bergenfeldt}\ \emph {et~al.}(2014)\citenamefont
  {Bergenfeldt}, \citenamefont {Samuelsson}, \citenamefont {Sothmann},
  \citenamefont {Flindt},\ and\ \citenamefont
  {Büttiker}}]{bergenfeldt_hybrid_2014}%
  \BibitemOpen
  \bibfield  {author} {\bibinfo {author} {\bibfnamefont {C.}~\bibnamefont
  {Bergenfeldt}}, \bibinfo {author} {\bibfnamefont {P.}~\bibnamefont
  {Samuelsson}}, \bibinfo {author} {\bibfnamefont {B.}~\bibnamefont
  {Sothmann}}, \bibinfo {author} {\bibfnamefont {C.}~\bibnamefont {Flindt}}, \
  and\ \bibinfo {author} {\bibfnamefont {M.}~\bibnamefont {Büttiker}},\ }\href
  {\doibase 10.1103/PhysRevLett.112.076803} {\bibfield  {journal} {\bibinfo
  {journal} {Phys. Rev. Lett.}\ }\textbf {\bibinfo {volume} {112}},\ \bibinfo
  {pages} {076803} (\bibinfo {year} {2014})}\BibitemShut {NoStop}%
\bibitem [{\citenamefont {Mazza}\ \emph {et~al.}(2014)\citenamefont {Mazza},
  \citenamefont {Bosisio}, \citenamefont {Benenti}, \citenamefont
  {Giovannetti}, \citenamefont {Fazio},\ and\ \citenamefont
  {Taddei}}]{mazza_thermoelectric_2014}%
  \BibitemOpen
  \bibfield  {author} {\bibinfo {author} {\bibfnamefont {F.}~\bibnamefont
  {Mazza}}, \bibinfo {author} {\bibfnamefont {R.}~\bibnamefont {Bosisio}},
  \bibinfo {author} {\bibfnamefont {G.}~\bibnamefont {Benenti}}, \bibinfo
  {author} {\bibfnamefont {V.}~\bibnamefont {Giovannetti}}, \bibinfo {author}
  {\bibfnamefont {R.}~\bibnamefont {Fazio}}, \ and\ \bibinfo {author}
  {\bibfnamefont {F.}~\bibnamefont {Taddei}},\ }\href {\doibase
  10.1088/1367-2630/16/8/085001} {\bibfield  {journal} {\bibinfo  {journal}
  {New J. Phys.}\ }\textbf {\bibinfo {volume} {16}},\ \bibinfo {pages} {085001}
  (\bibinfo {year} {2014})}\BibitemShut {NoStop}%
\bibitem [{\citenamefont {Mazza}\ \emph {et~al.}(2015)\citenamefont {Mazza},
  \citenamefont {Valentini}, \citenamefont {Bosisio}, \citenamefont {Benenti},
  \citenamefont {Giovannetti}, \citenamefont {Fazio},\ and\ \citenamefont
  {Taddei}}]{mazza_separation_2015}%
  \BibitemOpen
  \bibfield  {author} {\bibinfo {author} {\bibfnamefont {F.}~\bibnamefont
  {Mazza}}, \bibinfo {author} {\bibfnamefont {S.}~\bibnamefont {Valentini}},
  \bibinfo {author} {\bibfnamefont {R.}~\bibnamefont {Bosisio}}, \bibinfo
  {author} {\bibfnamefont {G.}~\bibnamefont {Benenti}}, \bibinfo {author}
  {\bibfnamefont {V.}~\bibnamefont {Giovannetti}}, \bibinfo {author}
  {\bibfnamefont {R.}~\bibnamefont {Fazio}}, \ and\ \bibinfo {author}
  {\bibfnamefont {F.}~\bibnamefont {Taddei}},\ }\href {\doibase
  10.1103/PhysRevB.91.245435} {\bibfield  {journal} {\bibinfo  {journal} {Phys.
  Rev. B}\ }\textbf {\bibinfo {volume} {91}},\ \bibinfo {pages} {245435}
  (\bibinfo {year} {2015})}\BibitemShut {NoStop}%
\bibitem [{\citenamefont {Sothmann}\ \emph {et~al.}(2015)\citenamefont
  {Sothmann}, \citenamefont {Sánchez},\ and\ \citenamefont
  {Jordan}}]{sothmann_thermoelectric_2015}%
  \BibitemOpen
  \bibfield  {author} {\bibinfo {author} {\bibfnamefont {B.}~\bibnamefont
  {Sothmann}}, \bibinfo {author} {\bibfnamefont {R.}~\bibnamefont {Sánchez}},
  \ and\ \bibinfo {author} {\bibfnamefont {A.~N.}\ \bibnamefont {Jordan}},\
  }\href {\doibase 10.1088/0957-4484/26/3/032001} {\bibfield  {journal}
  {\bibinfo  {journal} {Nanotechnology}\ }\textbf {\bibinfo {volume} {26}},\
  \bibinfo {pages} {032001} (\bibinfo {year} {2015})}\BibitemShut {NoStop}%
\bibitem [{\citenamefont {Hartmann}\ \emph {et~al.}(2015)\citenamefont
  {Hartmann}, \citenamefont {Pfeffer}, \citenamefont {Höfling}, \citenamefont
  {Kamp},\ and\ \citenamefont {Worschech}}]{hartmann_voltage_2015}%
  \BibitemOpen
  \bibfield  {author} {\bibinfo {author} {\bibfnamefont {F.}~\bibnamefont
  {Hartmann}}, \bibinfo {author} {\bibfnamefont {P.}~\bibnamefont {Pfeffer}},
  \bibinfo {author} {\bibfnamefont {S.}~\bibnamefont {Höfling}}, \bibinfo
  {author} {\bibfnamefont {M.}~\bibnamefont {Kamp}}, \ and\ \bibinfo {author}
  {\bibfnamefont {L.}~\bibnamefont {Worschech}},\ }\href {\doibase
  10.1103/PhysRevLett.114.146805} {\bibfield  {journal} {\bibinfo  {journal}
  {Phys. Rev. Lett.}\ }\textbf {\bibinfo {volume} {114}},\ \bibinfo {pages}
  {146805} (\bibinfo {year} {2015})}\BibitemShut {NoStop}%
\bibitem [{\citenamefont {Roche}\ \emph {et~al.}(2015)\citenamefont {Roche},
  \citenamefont {Roulleau}, \citenamefont {Jullien}, \citenamefont {Jompol},
  \citenamefont {Farrer}, \citenamefont {Ritchie},\ and\ \citenamefont
  {Glattli}}]{roche_harvesting_2015}%
  \BibitemOpen
  \bibfield  {author} {\bibinfo {author} {\bibfnamefont {B.}~\bibnamefont
  {Roche}}, \bibinfo {author} {\bibfnamefont {P.}~\bibnamefont {Roulleau}},
  \bibinfo {author} {\bibfnamefont {T.}~\bibnamefont {Jullien}}, \bibinfo
  {author} {\bibfnamefont {Y.}~\bibnamefont {Jompol}}, \bibinfo {author}
  {\bibfnamefont {I.}~\bibnamefont {Farrer}}, \bibinfo {author} {\bibfnamefont
  {D.~A.}\ \bibnamefont {Ritchie}}, \ and\ \bibinfo {author} {\bibfnamefont
  {D.~C.}\ \bibnamefont {Glattli}},\ }\href {\doibase 10.1038/ncomms7738}
  {\bibfield  {journal} {\bibinfo  {journal} {Nat. Commun.}\ }\textbf {\bibinfo
  {volume} {6}} (\bibinfo {year} {2015}),\ 10.1038/ncomms7738}\BibitemShut
  {NoStop}%
\bibitem [{\citenamefont {Thierschmann}\ \emph {et~al.}(2015)\citenamefont
  {Thierschmann}, \citenamefont {Sánchez}, \citenamefont {Sothmann},
  \citenamefont {Arnold}, \citenamefont {Heyn}, \citenamefont {Hansen},
  \citenamefont {Buhmann},\ and\ \citenamefont
  {Molenkamp}}]{thierschmann_three-terminal_2015}%
  \BibitemOpen
  \bibfield  {author} {\bibinfo {author} {\bibfnamefont {H.}~\bibnamefont
  {Thierschmann}}, \bibinfo {author} {\bibfnamefont {R.}~\bibnamefont
  {Sánchez}}, \bibinfo {author} {\bibfnamefont {B.}~\bibnamefont {Sothmann}},
  \bibinfo {author} {\bibfnamefont {F.}~\bibnamefont {Arnold}}, \bibinfo
  {author} {\bibfnamefont {C.}~\bibnamefont {Heyn}}, \bibinfo {author}
  {\bibfnamefont {W.}~\bibnamefont {Hansen}}, \bibinfo {author} {\bibfnamefont
  {H.}~\bibnamefont {Buhmann}}, \ and\ \bibinfo {author} {\bibfnamefont
  {L.~W.}\ \bibnamefont {Molenkamp}},\ }\href {\doibase 10.1038/nnano.2015.176}
  {\bibfield  {journal} {\bibinfo  {journal} {Nature Nanotech.}\ }\textbf
  {\bibinfo {volume} {10}},\ \bibinfo {pages} {854} (\bibinfo {year}
  {2015})}\BibitemShut {NoStop}%
\bibitem [{\citenamefont {Scully}\ \emph {et~al.}(2003)\citenamefont {Scully},
  \citenamefont {Zubairy}, \citenamefont {Agarwal},\ and\ \citenamefont
  {Walther}}]{scully_extracting_2003}%
  \BibitemOpen
  \bibfield  {author} {\bibinfo {author} {\bibfnamefont {M.~O.}\ \bibnamefont
  {Scully}}, \bibinfo {author} {\bibfnamefont {M.~S.}\ \bibnamefont {Zubairy}},
  \bibinfo {author} {\bibfnamefont {G.~S.}\ \bibnamefont {Agarwal}}, \ and\
  \bibinfo {author} {\bibfnamefont {H.}~\bibnamefont {Walther}},\ }\href
  {\doibase 10.1126/science.1078955} {\bibfield  {journal} {\bibinfo  {journal}
  {Science}\ }\textbf {\bibinfo {volume} {299}},\ \bibinfo {pages} {862}
  (\bibinfo {year} {2003})}\BibitemShut {NoStop}%
\bibitem [{\citenamefont {Roßnagel}\ \emph {et~al.}(2014)\citenamefont
  {Roßnagel}, \citenamefont {Abah}, \citenamefont {Schmidt-Kaler},
  \citenamefont {Singer},\ and\ \citenamefont
  {Lutz}}]{rosnagel_nanoscale_2014}%
  \BibitemOpen
  \bibfield  {author} {\bibinfo {author} {\bibfnamefont {J.}~\bibnamefont
  {Roßnagel}}, \bibinfo {author} {\bibfnamefont {O.}~\bibnamefont {Abah}},
  \bibinfo {author} {\bibfnamefont {F.}~\bibnamefont {Schmidt-Kaler}}, \bibinfo
  {author} {\bibfnamefont {K.}~\bibnamefont {Singer}}, \ and\ \bibinfo {author}
  {\bibfnamefont {E.}~\bibnamefont {Lutz}},\ }\href {\doibase
  10.1103/PhysRevLett.112.030602} {\bibfield  {journal} {\bibinfo  {journal}
  {Phys. Rev. Lett.}\ }\textbf {\bibinfo {volume} {112}},\ \bibinfo {pages}
  {030602} (\bibinfo {year} {2014})}\BibitemShut {NoStop}%
\bibitem [{\citenamefont {Neder}\ \emph
  {et~al.}(2007{\natexlab{a}})\citenamefont {Neder}, \citenamefont {Ofek},
  \citenamefont {Chung}, \citenamefont {Heiblum}, \citenamefont {Mahalu},\ and\
  \citenamefont {Umansky}}]{neder_interference_2007}%
  \BibitemOpen
  \bibfield  {author} {\bibinfo {author} {\bibfnamefont {I.}~\bibnamefont
  {Neder}}, \bibinfo {author} {\bibfnamefont {N.}~\bibnamefont {Ofek}},
  \bibinfo {author} {\bibfnamefont {Y.}~\bibnamefont {Chung}}, \bibinfo
  {author} {\bibfnamefont {M.}~\bibnamefont {Heiblum}}, \bibinfo {author}
  {\bibfnamefont {D.}~\bibnamefont {Mahalu}}, \ and\ \bibinfo {author}
  {\bibfnamefont {V.}~\bibnamefont {Umansky}},\ }\href {\doibase
  10.1038/nature05955} {\bibfield  {journal} {\bibinfo  {journal} {Nature}\
  }\textbf {\bibinfo {volume} {448}},\ \bibinfo {pages} {333} (\bibinfo {year}
  {2007}{\natexlab{a}})}\BibitemShut {NoStop}%
\bibitem [{\citenamefont {Sothmann}\ \emph {et~al.}(2014)\citenamefont
  {Sothmann}, \citenamefont {Sánchez},\ and\ \citenamefont
  {Jordan}}]{sothmann_quantum_2014}%
  \BibitemOpen
  \bibfield  {author} {\bibinfo {author} {\bibfnamefont {B.}~\bibnamefont
  {Sothmann}}, \bibinfo {author} {\bibfnamefont {R.}~\bibnamefont {Sánchez}},
  \ and\ \bibinfo {author} {\bibfnamefont {A.~N.}\ \bibnamefont {Jordan}},\
  }\href {\doibase 10.1209/0295-5075/107/47003} {\bibfield  {journal} {\bibinfo
   {journal} {Europhys. Lett.}\ }\textbf {\bibinfo {volume} {107}},\ \bibinfo
  {pages} {47003} (\bibinfo {year} {2014})}\BibitemShut {NoStop}%
\bibitem [{\citenamefont {López}\ \emph {et~al.}(2014)\citenamefont {López},
  \citenamefont {Hwang},\ and\ \citenamefont
  {Sánchez}}]{lopez_thermoelectric_2014}%
  \BibitemOpen
  \bibfield  {author} {\bibinfo {author} {\bibfnamefont {R.}~\bibnamefont
  {López}}, \bibinfo {author} {\bibfnamefont {S.-Y.}\ \bibnamefont {Hwang}}, \
  and\ \bibinfo {author} {\bibfnamefont {D.}~\bibnamefont {Sánchez}},\ }\href
  {\doibase 10.1088/1742-6596/568/5/052016} {\bibfield  {journal} {\bibinfo
  {journal} {J. Phys.: Conf. Ser.}\ }\textbf {\bibinfo {volume} {568}},\
  \bibinfo {pages} {052016} (\bibinfo {year} {2014})}\BibitemShut {NoStop}%
\bibitem [{\citenamefont {Sánchez}\ \emph
  {et~al.}(2015{\natexlab{a}})\citenamefont {Sánchez}, \citenamefont
  {Sothmann},\ and\ \citenamefont {Jordan}}]{sanchez_chiral_2015}%
  \BibitemOpen
  \bibfield  {author} {\bibinfo {author} {\bibfnamefont {R.}~\bibnamefont
  {Sánchez}}, \bibinfo {author} {\bibfnamefont {B.}~\bibnamefont {Sothmann}},
  \ and\ \bibinfo {author} {\bibfnamefont {A.~N.}\ \bibnamefont {Jordan}},\
  }\href {\doibase 10.1103/PhysRevLett.114.146801} {\bibfield  {journal}
  {\bibinfo  {journal} {Phys. Rev. Lett.}\ }\textbf {\bibinfo {volume} {114}},\
  \bibinfo {pages} {146801} (\bibinfo {year} {2015}{\natexlab{a}})}\BibitemShut
  {NoStop}%
\bibitem [{\citenamefont {Sánchez}\ \emph
  {et~al.}(2015{\natexlab{b}})\citenamefont {Sánchez}, \citenamefont
  {Sothmann},\ and\ \citenamefont {Jordan}}]{sanchez_heat_2015}%
  \BibitemOpen
  \bibfield  {author} {\bibinfo {author} {\bibfnamefont {R.}~\bibnamefont
  {Sánchez}}, \bibinfo {author} {\bibfnamefont {B.}~\bibnamefont {Sothmann}},
  \ and\ \bibinfo {author} {\bibfnamefont {A.~N.}\ \bibnamefont {Jordan}},\
  }\href {\doibase 10.1088/1367-2630/17/7/075006} {\bibfield  {journal}
  {\bibinfo  {journal} {New J. Phys.}\ }\textbf {\bibinfo {volume} {17}},\
  \bibinfo {pages} {075006} (\bibinfo {year} {2015}{\natexlab{b}})}\BibitemShut
  {NoStop}%
\bibitem [{\citenamefont {Sánchez}\ \emph {et~al.}(2016)\citenamefont
  {Sánchez}, \citenamefont {Sothmann},\ and\ \citenamefont
  {Jordan}}]{sanchez_effect_2016}%
  \BibitemOpen
  \bibfield  {author} {\bibinfo {author} {\bibfnamefont {R.}~\bibnamefont
  {Sánchez}}, \bibinfo {author} {\bibfnamefont {B.}~\bibnamefont {Sothmann}},
  \ and\ \bibinfo {author} {\bibfnamefont {A.~N.}\ \bibnamefont {Jordan}},\
  }\href {\doibase 10.1016/j.physe.2015.09.004} {\bibfield  {journal} {\bibinfo
   {journal} {Physica E}\ }\textbf {\bibinfo {volume} {75}},\ \bibinfo {pages}
  {86} (\bibinfo {year} {2016})}\BibitemShut {NoStop}%
\bibitem [{\citenamefont {Hofer}\ and\ \citenamefont
  {Sothmann}(2015)}]{hofer_quantum_2015}%
  \BibitemOpen
  \bibfield  {author} {\bibinfo {author} {\bibfnamefont {P.~P.}\ \bibnamefont
  {Hofer}}\ and\ \bibinfo {author} {\bibfnamefont {B.}~\bibnamefont
  {Sothmann}},\ }\href {\doibase 10.1103/PhysRevB.91.195406} {\bibfield
  {journal} {\bibinfo  {journal} {Phys. Rev. B}\ }\textbf {\bibinfo {volume}
  {91}},\ \bibinfo {pages} {195406} (\bibinfo {year} {2015})}\BibitemShut
  {NoStop}%
\bibitem [{\citenamefont {Benenti}\ \emph {et~al.}(2011)\citenamefont
  {Benenti}, \citenamefont {Saito},\ and\ \citenamefont
  {Casati}}]{benenti_thermodynamic_2011}%
  \BibitemOpen
  \bibfield  {author} {\bibinfo {author} {\bibfnamefont {G.}~\bibnamefont
  {Benenti}}, \bibinfo {author} {\bibfnamefont {K.}~\bibnamefont {Saito}}, \
  and\ \bibinfo {author} {\bibfnamefont {G.}~\bibnamefont {Casati}},\ }\href
  {\doibase 10.1103/PhysRevLett.106.230602} {\bibfield  {journal} {\bibinfo
  {journal} {Phys. Rev. Lett.}\ }\textbf {\bibinfo {volume} {106}},\ \bibinfo
  {pages} {230602} (\bibinfo {year} {2011})}\BibitemShut {NoStop}%
\bibitem [{\citenamefont {Saito}\ \emph {et~al.}(2011)\citenamefont {Saito},
  \citenamefont {Benenti}, \citenamefont {Casati},\ and\ \citenamefont
  {Prosen}}]{saito_thermopower_2011}%
  \BibitemOpen
  \bibfield  {author} {\bibinfo {author} {\bibfnamefont {K.}~\bibnamefont
  {Saito}}, \bibinfo {author} {\bibfnamefont {G.}~\bibnamefont {Benenti}},
  \bibinfo {author} {\bibfnamefont {G.}~\bibnamefont {Casati}}, \ and\ \bibinfo
  {author} {\bibfnamefont {T.}~\bibnamefont {Prosen}},\ }\href {\doibase
  10.1103/PhysRevB.84.201306} {\bibfield  {journal} {\bibinfo  {journal} {Phys.
  Rev. B}\ }\textbf {\bibinfo {volume} {84}},\ \bibinfo {pages} {201306}
  (\bibinfo {year} {2011})}\BibitemShut {NoStop}%
\bibitem [{\citenamefont {Sánchez}\ and\ \citenamefont
  {Serra}(2011)}]{sanchez_thermoelectric_2011}%
  \BibitemOpen
  \bibfield  {author} {\bibinfo {author} {\bibfnamefont {D.}~\bibnamefont
  {Sánchez}}\ and\ \bibinfo {author} {\bibfnamefont {L.}~\bibnamefont
  {Serra}},\ }\href {\doibase 10.1103/PhysRevB.84.201307} {\bibfield  {journal}
  {\bibinfo  {journal} {Phys. Rev. B}\ }\textbf {\bibinfo {volume} {84}},\
  \bibinfo {pages} {201307} (\bibinfo {year} {2011})}\BibitemShut {NoStop}%
\bibitem [{\citenamefont {Brandner}\ \emph {et~al.}(2013)\citenamefont
  {Brandner}, \citenamefont {Saito},\ and\ \citenamefont
  {Seifert}}]{brandner_strong_2013}%
  \BibitemOpen
  \bibfield  {author} {\bibinfo {author} {\bibfnamefont {K.}~\bibnamefont
  {Brandner}}, \bibinfo {author} {\bibfnamefont {K.}~\bibnamefont {Saito}}, \
  and\ \bibinfo {author} {\bibfnamefont {U.}~\bibnamefont {Seifert}},\ }\href
  {\doibase 10.1103/PhysRevLett.110.070603} {\bibfield  {journal} {\bibinfo
  {journal} {Phys. Rev. Lett.}\ }\textbf {\bibinfo {volume} {110}},\ \bibinfo
  {pages} {070603} (\bibinfo {year} {2013})}\BibitemShut {NoStop}%
\bibitem [{\citenamefont {Brandner}\ and\ \citenamefont
  {Seifert}(2013)}]{brandner_multi-terminal_2013}%
  \BibitemOpen
  \bibfield  {author} {\bibinfo {author} {\bibfnamefont {K.}~\bibnamefont
  {Brandner}}\ and\ \bibinfo {author} {\bibfnamefont {U.}~\bibnamefont
  {Seifert}},\ }\href {\doibase 10.1088/1367-2630/15/10/105003} {\bibfield
  {journal} {\bibinfo  {journal} {New J. Phys.}\ }\textbf {\bibinfo {volume}
  {15}},\ \bibinfo {pages} {105003} (\bibinfo {year} {2013})}\BibitemShut
  {NoStop}%
\bibitem [{\citenamefont {Granger}\ \emph {et~al.}(2009)\citenamefont
  {Granger}, \citenamefont {Eisenstein},\ and\ \citenamefont
  {Reno}}]{granger_observation_2009}%
  \BibitemOpen
  \bibfield  {author} {\bibinfo {author} {\bibfnamefont {G.}~\bibnamefont
  {Granger}}, \bibinfo {author} {\bibfnamefont {J.~P.}\ \bibnamefont
  {Eisenstein}}, \ and\ \bibinfo {author} {\bibfnamefont {J.~L.}\ \bibnamefont
  {Reno}},\ }\href {\doibase 10.1103/PhysRevLett.102.086803} {\bibfield
  {journal} {\bibinfo  {journal} {Phys. Rev. Lett.}\ }\textbf {\bibinfo
  {volume} {102}},\ \bibinfo {pages} {086803} (\bibinfo {year}
  {2009})}\BibitemShut {NoStop}%
\bibitem [{\citenamefont {Altimiras}\ \emph {et~al.}(2010)\citenamefont
  {Altimiras}, \citenamefont {le~Sueur}, \citenamefont {Gennser}, \citenamefont
  {Cavanna}, \citenamefont {Mailly},\ and\ \citenamefont
  {Pierre}}]{altimiras_non-equilibrium_2010}%
  \BibitemOpen
  \bibfield  {author} {\bibinfo {author} {\bibfnamefont {C.}~\bibnamefont
  {Altimiras}}, \bibinfo {author} {\bibfnamefont {H.}~\bibnamefont {le~Sueur}},
  \bibinfo {author} {\bibfnamefont {U.}~\bibnamefont {Gennser}}, \bibinfo
  {author} {\bibfnamefont {A.}~\bibnamefont {Cavanna}}, \bibinfo {author}
  {\bibfnamefont {D.}~\bibnamefont {Mailly}}, \ and\ \bibinfo {author}
  {\bibfnamefont {F.}~\bibnamefont {Pierre}},\ }\href {\doibase
  10.1038/nphys1429} {\bibfield  {journal} {\bibinfo  {journal} {Nat. Phys.}\
  }\textbf {\bibinfo {volume} {6}},\ \bibinfo {pages} {34} (\bibinfo {year}
  {2010})}\BibitemShut {NoStop}%
\bibitem [{\citenamefont {Nam}\ \emph {et~al.}(2013)\citenamefont {Nam},
  \citenamefont {Hwang},\ and\ \citenamefont {Lee}}]{nam_thermoelectric_2013}%
  \BibitemOpen
  \bibfield  {author} {\bibinfo {author} {\bibfnamefont {S.-G.}\ \bibnamefont
  {Nam}}, \bibinfo {author} {\bibfnamefont {E.~H.}\ \bibnamefont {Hwang}}, \
  and\ \bibinfo {author} {\bibfnamefont {H.-J.}\ \bibnamefont {Lee}},\ }\href
  {\doibase 10.1103/PhysRevLett.110.226801} {\bibfield  {journal} {\bibinfo
  {journal} {Phys. Rev. Lett.}\ }\textbf {\bibinfo {volume} {110}},\ \bibinfo
  {pages} {226801} (\bibinfo {year} {2013})}\BibitemShut {NoStop}%
\bibitem [{\citenamefont {Jezouin}\ \emph {et~al.}(2013)\citenamefont
  {Jezouin}, \citenamefont {Parmentier}, \citenamefont {Anthore}, \citenamefont
  {Gennser}, \citenamefont {Cavanna}, \citenamefont {Jin},\ and\ \citenamefont
  {Pierre}}]{jezouin_quantum_2013}%
  \BibitemOpen
  \bibfield  {author} {\bibinfo {author} {\bibfnamefont {S.}~\bibnamefont
  {Jezouin}}, \bibinfo {author} {\bibfnamefont {F.~D.}\ \bibnamefont
  {Parmentier}}, \bibinfo {author} {\bibfnamefont {A.}~\bibnamefont {Anthore}},
  \bibinfo {author} {\bibfnamefont {U.}~\bibnamefont {Gennser}}, \bibinfo
  {author} {\bibfnamefont {A.}~\bibnamefont {Cavanna}}, \bibinfo {author}
  {\bibfnamefont {Y.}~\bibnamefont {Jin}}, \ and\ \bibinfo {author}
  {\bibfnamefont {F.}~\bibnamefont {Pierre}},\ }\href {\doibase
  10.1126/science.1241912} {\bibfield  {journal} {\bibinfo  {journal}
  {Science}\ }\textbf {\bibinfo {volume} {342}},\ \bibinfo {pages} {601}
  (\bibinfo {year} {2013})}\BibitemShut {NoStop}%
\bibitem [{\citenamefont {Levkivskyi}\ and\ \citenamefont
  {Sukhorukov}(2012)}]{levkivskyi_energy_2012}%
  \BibitemOpen
  \bibfield  {author} {\bibinfo {author} {\bibfnamefont {I.~P.}\ \bibnamefont
  {Levkivskyi}}\ and\ \bibinfo {author} {\bibfnamefont {E.~V.}\ \bibnamefont
  {Sukhorukov}},\ }\href {\doibase 10.1103/PhysRevB.85.075309} {\bibfield
  {journal} {\bibinfo  {journal} {Phys. Rev. B}\ }\textbf {\bibinfo {volume}
  {85}},\ \bibinfo {pages} {075309} (\bibinfo {year} {2012})}\BibitemShut
  {NoStop}%
\bibitem [{\citenamefont {Aita}\ \emph {et~al.}(2013)\citenamefont {Aita},
  \citenamefont {Arrachea}, \citenamefont {Naón},\ and\ \citenamefont
  {Fradkin}}]{aita_heat_2013}%
  \BibitemOpen
  \bibfield  {author} {\bibinfo {author} {\bibfnamefont {H.}~\bibnamefont
  {Aita}}, \bibinfo {author} {\bibfnamefont {L.}~\bibnamefont {Arrachea}},
  \bibinfo {author} {\bibfnamefont {C.}~\bibnamefont {Naón}}, \ and\ \bibinfo
  {author} {\bibfnamefont {E.}~\bibnamefont {Fradkin}},\ }\href {\doibase
  10.1103/PhysRevB.88.085122} {\bibfield  {journal} {\bibinfo  {journal} {Phys.
  Rev. B}\ }\textbf {\bibinfo {volume} {88}},\ \bibinfo {pages} {085122}
  (\bibinfo {year} {2013})}\BibitemShut {NoStop}%
\bibitem [{\citenamefont {Vannucci}\ \emph {et~al.}(2015)\citenamefont
  {Vannucci}, \citenamefont {Ronetti}, \citenamefont {Dolcetto}, \citenamefont
  {Carrega},\ and\ \citenamefont
  {Sassetti}}]{vannucci_interference-induced_2015}%
  \BibitemOpen
  \bibfield  {author} {\bibinfo {author} {\bibfnamefont {L.}~\bibnamefont
  {Vannucci}}, \bibinfo {author} {\bibfnamefont {F.}~\bibnamefont {Ronetti}},
  \bibinfo {author} {\bibfnamefont {G.}~\bibnamefont {Dolcetto}}, \bibinfo
  {author} {\bibfnamefont {M.}~\bibnamefont {Carrega}}, \ and\ \bibinfo
  {author} {\bibfnamefont {M.}~\bibnamefont {Sassetti}},\ }\href {\doibase
  10.1103/PhysRevB.92.075446} {\bibfield  {journal} {\bibinfo  {journal} {Phys.
  Rev. B}\ }\textbf {\bibinfo {volume} {92}},\ \bibinfo {pages} {075446}
  (\bibinfo {year} {2015})}\BibitemShut {NoStop}%
\bibitem [{\citenamefont {Goldstein}\ and\ \citenamefont
  {Gefen}(2016)}]{goldstein_suppression_2016}%
  \BibitemOpen
  \bibfield  {author} {\bibinfo {author} {\bibfnamefont {M.}~\bibnamefont
  {Goldstein}}\ and\ \bibinfo {author} {\bibfnamefont {Y.}~\bibnamefont
  {Gefen}},\ }\href {\doibase 10.1103/PhysRevLett.117.276804} {\bibfield
  {journal} {\bibinfo  {journal} {Phys. Rev. Lett.}\ }\textbf {\bibinfo
  {volume} {117}},\ \bibinfo {pages} {276804} (\bibinfo {year}
  {2016})}\BibitemShut {NoStop}%
\bibitem [{\citenamefont {Jordan}\ \emph {et~al.}(2013)\citenamefont {Jordan},
  \citenamefont {Sothmann}, \citenamefont {Sánchez},\ and\ \citenamefont
  {Büttiker}}]{jordan_powerful_2013}%
  \BibitemOpen
  \bibfield  {author} {\bibinfo {author} {\bibfnamefont {A.~N.}\ \bibnamefont
  {Jordan}}, \bibinfo {author} {\bibfnamefont {B.}~\bibnamefont {Sothmann}},
  \bibinfo {author} {\bibfnamefont {R.}~\bibnamefont {Sánchez}}, \ and\
  \bibinfo {author} {\bibfnamefont {M.}~\bibnamefont {Büttiker}},\ }\href
  {\doibase 10.1103/PhysRevB.87.075312} {\bibfield  {journal} {\bibinfo
  {journal} {Phys. Rev. B}\ }\textbf {\bibinfo {volume} {87}},\ \bibinfo
  {pages} {075312} (\bibinfo {year} {2013})}\BibitemShut {NoStop}%
\bibitem [{\citenamefont {Whitney}(2014)}]{whitney_most_2014}%
  \BibitemOpen
  \bibfield  {author} {\bibinfo {author} {\bibfnamefont {R.~S.}\ \bibnamefont
  {Whitney}},\ }\href {\doibase 10.1103/PhysRevLett.112.130601} {\bibfield
  {journal} {\bibinfo  {journal} {Phys. Rev. Lett.}\ }\textbf {\bibinfo
  {volume} {112}},\ \bibinfo {pages} {130601} (\bibinfo {year}
  {2014})}\BibitemShut {NoStop}%
\bibitem [{\citenamefont {Whitney}(2015)}]{whitney_finding_2015}%
  \BibitemOpen
  \bibfield  {author} {\bibinfo {author} {\bibfnamefont {R.~S.}\ \bibnamefont
  {Whitney}},\ }\href {\doibase 10.1103/PhysRevB.91.115425} {\bibfield
  {journal} {\bibinfo  {journal} {Phys. Rev. B}\ }\textbf {\bibinfo {volume}
  {91}},\ \bibinfo {pages} {115425} (\bibinfo {year} {2015})}\BibitemShut
  {NoStop}%
\bibitem [{\citenamefont {Büttiker}\ \emph {et~al.}(1993)\citenamefont
  {Büttiker}, \citenamefont {Thomas},\ and\ \citenamefont
  {Prêtre}}]{buttiker_mesoscopic_1993}%
  \BibitemOpen
  \bibfield  {author} {\bibinfo {author} {\bibfnamefont {M.}~\bibnamefont
  {Büttiker}}, \bibinfo {author} {\bibfnamefont {H.}~\bibnamefont {Thomas}}, \
  and\ \bibinfo {author} {\bibfnamefont {A.}~\bibnamefont {Prêtre}},\ }\href
  {\doibase 10.1016/0375-9601(93)91193-9} {\bibfield  {journal} {\bibinfo
  {journal} {Phys. Lett. A}\ }\textbf {\bibinfo {volume} {180}},\ \bibinfo
  {pages} {364} (\bibinfo {year} {1993})}\BibitemShut {NoStop}%
\bibitem [{\citenamefont {Gabelli}\ \emph {et~al.}(2006)\citenamefont
  {Gabelli}, \citenamefont {Fève}, \citenamefont {Berroir}, \citenamefont
  {Plaçais}, \citenamefont {Cavanna}, \citenamefont {Etienne}, \citenamefont
  {Jin},\ and\ \citenamefont {Glattli}}]{gabelli_violation_2006}%
  \BibitemOpen
  \bibfield  {author} {\bibinfo {author} {\bibfnamefont {J.}~\bibnamefont
  {Gabelli}}, \bibinfo {author} {\bibfnamefont {G.}~\bibnamefont {Fève}},
  \bibinfo {author} {\bibfnamefont {J.-M.}\ \bibnamefont {Berroir}}, \bibinfo
  {author} {\bibfnamefont {B.}~\bibnamefont {Plaçais}}, \bibinfo {author}
  {\bibfnamefont {A.}~\bibnamefont {Cavanna}}, \bibinfo {author} {\bibfnamefont
  {B.}~\bibnamefont {Etienne}}, \bibinfo {author} {\bibfnamefont
  {Y.}~\bibnamefont {Jin}}, \ and\ \bibinfo {author} {\bibfnamefont {D.~C.}\
  \bibnamefont {Glattli}},\ }\href {\doibase 10.1126/science.1126940}
  {\bibfield  {journal} {\bibinfo  {journal} {Science}\ }\textbf {\bibinfo
  {volume} {313}},\ \bibinfo {pages} {499 } (\bibinfo {year}
  {2006})}\BibitemShut {NoStop}%
\bibitem [{\citenamefont {Fève}\ \emph {et~al.}(2007)\citenamefont {Fève},
  \citenamefont {Mahé}, \citenamefont {Berroir}, \citenamefont {Kontos},
  \citenamefont {Plaçais}, \citenamefont {Glattli}, \citenamefont {Cavanna},
  \citenamefont {Etienne},\ and\ \citenamefont {Jin}}]{feve_-demand_2007}%
  \BibitemOpen
  \bibfield  {author} {\bibinfo {author} {\bibfnamefont {G.}~\bibnamefont
  {Fève}}, \bibinfo {author} {\bibfnamefont {A.}~\bibnamefont {Mahé}},
  \bibinfo {author} {\bibfnamefont {J.-M.}\ \bibnamefont {Berroir}}, \bibinfo
  {author} {\bibfnamefont {T.}~\bibnamefont {Kontos}}, \bibinfo {author}
  {\bibfnamefont {B.}~\bibnamefont {Plaçais}}, \bibinfo {author}
  {\bibfnamefont {D.~C.}\ \bibnamefont {Glattli}}, \bibinfo {author}
  {\bibfnamefont {A.}~\bibnamefont {Cavanna}}, \bibinfo {author} {\bibfnamefont
  {B.}~\bibnamefont {Etienne}}, \ and\ \bibinfo {author} {\bibfnamefont
  {Y.}~\bibnamefont {Jin}},\ }\href {\doibase 10.1126/science.1141243}
  {\bibfield  {journal} {\bibinfo  {journal} {Science}\ }\textbf {\bibinfo
  {volume} {316}},\ \bibinfo {pages} {1169} (\bibinfo {year}
  {2007})}\BibitemShut {NoStop}%
\bibitem [{\citenamefont {Ji}\ \emph {et~al.}(2003)\citenamefont {Ji},
  \citenamefont {Chung}, \citenamefont {Sprinzak}, \citenamefont {Heiblum},
  \citenamefont {Mahalu},\ and\ \citenamefont
  {Shtrikman}}]{ji_electronic_2003}%
  \BibitemOpen
  \bibfield  {author} {\bibinfo {author} {\bibfnamefont {Y.}~\bibnamefont
  {Ji}}, \bibinfo {author} {\bibfnamefont {Y.}~\bibnamefont {Chung}}, \bibinfo
  {author} {\bibfnamefont {D.}~\bibnamefont {Sprinzak}}, \bibinfo {author}
  {\bibfnamefont {M.}~\bibnamefont {Heiblum}}, \bibinfo {author} {\bibfnamefont
  {D.}~\bibnamefont {Mahalu}}, \ and\ \bibinfo {author} {\bibfnamefont
  {H.}~\bibnamefont {Shtrikman}},\ }\href {\doibase 10.1038/nature01503}
  {\bibfield  {journal} {\bibinfo  {journal} {Nature}\ }\textbf {\bibinfo
  {volume} {422}},\ \bibinfo {pages} {415} (\bibinfo {year}
  {2003})}\BibitemShut {NoStop}%
\bibitem [{\citenamefont {Ngo~Dinh}\ \emph {et~al.}(2013)\citenamefont
  {Ngo~Dinh}, \citenamefont {Bagrets},\ and\ \citenamefont
  {Mirlin}}]{ngo_dinh_analytically_2013}%
  \BibitemOpen
  \bibfield  {author} {\bibinfo {author} {\bibfnamefont {S.}~\bibnamefont
  {Ngo~Dinh}}, \bibinfo {author} {\bibfnamefont {D.~A.}\ \bibnamefont
  {Bagrets}}, \ and\ \bibinfo {author} {\bibfnamefont {A.~D.}\ \bibnamefont
  {Mirlin}},\ }\href {\doibase 10.1103/PhysRevB.87.195433} {\bibfield
  {journal} {\bibinfo  {journal} {Phys. Rev. B}\ }\textbf {\bibinfo {volume}
  {87}},\ \bibinfo {pages} {195433} (\bibinfo {year} {2013})}\BibitemShut
  {NoStop}%
\bibitem [{\citenamefont {Oppenheim}\ and\ \citenamefont
  {Schafer}(2009)}]{oppenheim_discrete-time_2009}%
  \BibitemOpen
  \bibfield  {author} {\bibinfo {author} {\bibfnamefont {A.~V.}\ \bibnamefont
  {Oppenheim}}\ and\ \bibinfo {author} {\bibfnamefont {R.~W.}\ \bibnamefont
  {Schafer}},\ }\href@noop {} {{\emph {\bibinfo
  {title} {Discrete-time signal processing}}}},\ \bibinfo {edition} {3rd}\ ed.\
  (\bibinfo  {publisher} {Pearson},\ \bibinfo {address} {Upper Saddle River},\
  \bibinfo {year} {2009})\BibitemShut {NoStop}%
\bibitem [{\citenamefont {Büttiker}(1988)}]{buttiker_absence_1988}%
  \BibitemOpen
  \bibfield  {author} {\bibinfo {author} {\bibfnamefont {M.}~\bibnamefont
  {Büttiker}},\ }\href {\doibase 10.1103/PhysRevB.38.9375} {\bibfield
  {journal} {\bibinfo  {journal} {Phys. Rev. B}\ }\textbf {\bibinfo {volume}
  {38}},\ \bibinfo {pages} {9375} (\bibinfo {year} {1988})}\BibitemShut
  {NoStop}%
\bibitem [{\citenamefont {Butcher}(1990)}]{butcher_thermal_1990}%
  \BibitemOpen
  \bibfield  {author} {\bibinfo {author} {\bibfnamefont {P.~N.}\ \bibnamefont
  {Butcher}},\ }\href {\doibase 10.1088/0953-8984/2/22/008} {\bibfield
  {journal} {\bibinfo  {journal} {J. Phys.: Condens. Matter}\ }\textbf
  {\bibinfo {volume} {2}},\ \bibinfo {pages} {4869} (\bibinfo {year}
  {1990})}\BibitemShut {NoStop}%
\bibitem [{\citenamefont {Neder}\ \emph
  {et~al.}(2007{\natexlab{b}})\citenamefont {Neder}, \citenamefont {Marquardt},
  \citenamefont {Heiblum}, \citenamefont {Mahalu},\ and\ \citenamefont
  {Umansky}}]{neder_controlled_2007}%
  \BibitemOpen
  \bibfield  {author} {\bibinfo {author} {\bibfnamefont {I.}~\bibnamefont
  {Neder}}, \bibinfo {author} {\bibfnamefont {F.}~\bibnamefont {Marquardt}},
  \bibinfo {author} {\bibfnamefont {M.}~\bibnamefont {Heiblum}}, \bibinfo
  {author} {\bibfnamefont {D.}~\bibnamefont {Mahalu}}, \ and\ \bibinfo {author}
  {\bibfnamefont {V.}~\bibnamefont {Umansky}},\ }\href {\doibase
  10.1038/nphys627} {\bibfield  {journal} {\bibinfo  {journal} {Nat. Phys.}\
  }\textbf {\bibinfo {volume} {3}},\ \bibinfo {pages} {534} (\bibinfo {year}
  {2007}{\natexlab{b}})}\BibitemShut {NoStop}%
\bibitem [{\citenamefont {Neder}\ \emph
  {et~al.}(2007{\natexlab{c}})\citenamefont {Neder}, \citenamefont {Heiblum},
  \citenamefont {Mahalu},\ and\ \citenamefont
  {Umansky}}]{neder_entanglement_2007}%
  \BibitemOpen
  \bibfield  {author} {\bibinfo {author} {\bibfnamefont {I.}~\bibnamefont
  {Neder}}, \bibinfo {author} {\bibfnamefont {M.}~\bibnamefont {Heiblum}},
  \bibinfo {author} {\bibfnamefont {D.}~\bibnamefont {Mahalu}}, \ and\ \bibinfo
  {author} {\bibfnamefont {V.}~\bibnamefont {Umansky}},\ }\href {\doibase
  10.1103/PhysRevLett.98.036803} {\bibfield  {journal} {\bibinfo  {journal}
  {Phys. Rev. Lett.}\ }\textbf {\bibinfo {volume} {98}},\ \bibinfo {pages}
  {036803} (\bibinfo {year} {2007}{\natexlab{c}})}\BibitemShut {NoStop}%
\bibitem [{\citenamefont {Roulleau}\ \emph
  {et~al.}(2008{\natexlab{a}})\citenamefont {Roulleau}, \citenamefont
  {Portier}, \citenamefont {Roche}, \citenamefont {Cavanna}, \citenamefont
  {Faini}, \citenamefont {Gennser},\ and\ \citenamefont
  {Mailly}}]{roulleau_direct_2008}%
  \BibitemOpen
  \bibfield  {author} {\bibinfo {author} {\bibfnamefont {P.}~\bibnamefont
  {Roulleau}}, \bibinfo {author} {\bibfnamefont {F.}~\bibnamefont {Portier}},
  \bibinfo {author} {\bibfnamefont {P.}~\bibnamefont {Roche}}, \bibinfo
  {author} {\bibfnamefont {A.}~\bibnamefont {Cavanna}}, \bibinfo {author}
  {\bibfnamefont {G.}~\bibnamefont {Faini}}, \bibinfo {author} {\bibfnamefont
  {U.}~\bibnamefont {Gennser}}, \ and\ \bibinfo {author} {\bibfnamefont
  {D.}~\bibnamefont {Mailly}},\ }\href {\doibase
  10.1103/PhysRevLett.100.126802} {\bibfield  {journal} {\bibinfo  {journal}
  {Phys. Rev. Lett.}\ }\textbf {\bibinfo {volume} {100}},\ \bibinfo {pages}
  {126802} (\bibinfo {year} {2008}{\natexlab{a}})}\BibitemShut {NoStop}%
\bibitem [{\citenamefont {Roulleau}\ \emph
  {et~al.}(2008{\natexlab{b}})\citenamefont {Roulleau}, \citenamefont
  {Portier}, \citenamefont {Roche}, \citenamefont {Cavanna}, \citenamefont
  {Faini}, \citenamefont {Gennser},\ and\ \citenamefont
  {Mailly}}]{roulleau_noise_2008}%
  \BibitemOpen
  \bibfield  {author} {\bibinfo {author} {\bibfnamefont {P.}~\bibnamefont
  {Roulleau}}, \bibinfo {author} {\bibfnamefont {F.}~\bibnamefont {Portier}},
  \bibinfo {author} {\bibfnamefont {P.}~\bibnamefont {Roche}}, \bibinfo
  {author} {\bibfnamefont {A.}~\bibnamefont {Cavanna}}, \bibinfo {author}
  {\bibfnamefont {G.}~\bibnamefont {Faini}}, \bibinfo {author} {\bibfnamefont
  {U.}~\bibnamefont {Gennser}}, \ and\ \bibinfo {author} {\bibfnamefont
  {D.}~\bibnamefont {Mailly}},\ }\href {\doibase
  10.1103/PhysRevLett.101.186803} {\bibfield  {journal} {\bibinfo  {journal}
  {Phys. Rev. Lett.}\ }\textbf {\bibinfo {volume} {101}},\ \bibinfo {pages}
  {186803} (\bibinfo {year} {2008}{\natexlab{b}})}\BibitemShut {NoStop}%
\bibitem [{\citenamefont {Roulleau}\ \emph {et~al.}(2009)\citenamefont
  {Roulleau}, \citenamefont {Portier}, \citenamefont {Roche}, \citenamefont
  {Cavanna}, \citenamefont {Faini}, \citenamefont {Gennser},\ and\
  \citenamefont {Mailly}}]{roulleau_tuning_2009}%
  \BibitemOpen
  \bibfield  {author} {\bibinfo {author} {\bibfnamefont {P.}~\bibnamefont
  {Roulleau}}, \bibinfo {author} {\bibfnamefont {F.}~\bibnamefont {Portier}},
  \bibinfo {author} {\bibfnamefont {P.}~\bibnamefont {Roche}}, \bibinfo
  {author} {\bibfnamefont {A.}~\bibnamefont {Cavanna}}, \bibinfo {author}
  {\bibfnamefont {G.}~\bibnamefont {Faini}}, \bibinfo {author} {\bibfnamefont
  {U.}~\bibnamefont {Gennser}}, \ and\ \bibinfo {author} {\bibfnamefont
  {D.}~\bibnamefont {Mailly}},\ }\href {\doibase
  10.1103/PhysRevLett.102.236802} {\bibfield  {journal} {\bibinfo  {journal}
  {Phys. Rev. Lett.}\ }\textbf {\bibinfo {volume} {102}},\ \bibinfo {pages}
  {236802} (\bibinfo {year} {2009})}\BibitemShut {NoStop}%
\bibitem [{\citenamefont {Huynh}\ \emph {et~al.}(2012)\citenamefont {Huynh},
  \citenamefont {Portier}, \citenamefont {le~Sueur}, \citenamefont {Faini},
  \citenamefont {Gennser}, \citenamefont {Mailly}, \citenamefont {Pierre},
  \citenamefont {Wegscheider},\ and\ \citenamefont
  {Roche}}]{huynh_quantum_2012}%
  \BibitemOpen
  \bibfield  {author} {\bibinfo {author} {\bibfnamefont {P.-A.}\ \bibnamefont
  {Huynh}}, \bibinfo {author} {\bibfnamefont {F.}~\bibnamefont {Portier}},
  \bibinfo {author} {\bibfnamefont {H.}~\bibnamefont {le~Sueur}}, \bibinfo
  {author} {\bibfnamefont {G.}~\bibnamefont {Faini}}, \bibinfo {author}
  {\bibfnamefont {U.}~\bibnamefont {Gennser}}, \bibinfo {author} {\bibfnamefont
  {D.}~\bibnamefont {Mailly}}, \bibinfo {author} {\bibfnamefont
  {F.}~\bibnamefont {Pierre}}, \bibinfo {author} {\bibfnamefont
  {W.}~\bibnamefont {Wegscheider}}, \ and\ \bibinfo {author} {\bibfnamefont
  {P.}~\bibnamefont {Roche}},\ }\href {\doibase 10.1103/PhysRevLett.108.256802}
  {\bibfield  {journal} {\bibinfo  {journal} {Phys. Rev. Lett.}\ }\textbf
  {\bibinfo {volume} {108}},\ \bibinfo {pages} {256802} (\bibinfo {year}
  {2012})}\BibitemShut {NoStop}%
\bibitem [{\citenamefont {Helzel}\ \emph {et~al.}(2015)\citenamefont {Helzel},
  \citenamefont {Litvin}, \citenamefont {Levkivskyi}, \citenamefont
  {Sukhorukov}, \citenamefont {Wegscheider},\ and\ \citenamefont
  {Strunk}}]{helzel_counting_2015}%
  \BibitemOpen
  \bibfield  {author} {\bibinfo {author} {\bibfnamefont {A.}~\bibnamefont
  {Helzel}}, \bibinfo {author} {\bibfnamefont {L.~V.}\ \bibnamefont {Litvin}},
  \bibinfo {author} {\bibfnamefont {I.~P.}\ \bibnamefont {Levkivskyi}},
  \bibinfo {author} {\bibfnamefont {E.~V.}\ \bibnamefont {Sukhorukov}},
  \bibinfo {author} {\bibfnamefont {W.}~\bibnamefont {Wegscheider}}, \ and\
  \bibinfo {author} {\bibfnamefont {C.}~\bibnamefont {Strunk}},\ }\href
  {\doibase 10.1103/PhysRevB.91.245419} {\bibfield  {journal} {\bibinfo
  {journal} {Phys. Rev. B}\ }\textbf {\bibinfo {volume} {91}},\ \bibinfo
  {pages} {245419} (\bibinfo {year} {2015})}\BibitemShut {NoStop}%
\bibitem [{\citenamefont {Pilgram}\ \emph {et~al.}(2006)\citenamefont
  {Pilgram}, \citenamefont {Samuelsson}, \citenamefont {Förster},\ and\
  \citenamefont {Büttiker}}]{pilgram_full-counting_2006}%
  \BibitemOpen
  \bibfield  {author} {\bibinfo {author} {\bibfnamefont {S.}~\bibnamefont
  {Pilgram}}, \bibinfo {author} {\bibfnamefont {P.}~\bibnamefont {Samuelsson}},
  \bibinfo {author} {\bibfnamefont {H.}~\bibnamefont {Förster}}, \ and\
  \bibinfo {author} {\bibfnamefont {M.}~\bibnamefont {Büttiker}},\ }\href
  {\doibase 10.1103/PhysRevLett.97.066801} {\bibfield  {journal} {\bibinfo
  {journal} {Phys. Rev. Lett.}\ }\textbf {\bibinfo {volume} {97}},\ \bibinfo
  {pages} {066801} (\bibinfo {year} {2006})}\BibitemShut {NoStop}%
\bibitem [{\citenamefont {Förster}\ \emph {et~al.}(2007)\citenamefont
  {Förster}, \citenamefont {Samuelsson}, \citenamefont {Pilgram},\ and\
  \citenamefont {Büttiker}}]{forster_voltage_2007}%
  \BibitemOpen
  \bibfield  {author} {\bibinfo {author} {\bibfnamefont {H.}~\bibnamefont
  {Förster}}, \bibinfo {author} {\bibfnamefont {P.}~\bibnamefont
  {Samuelsson}}, \bibinfo {author} {\bibfnamefont {S.}~\bibnamefont {Pilgram}},
  \ and\ \bibinfo {author} {\bibfnamefont {M.}~\bibnamefont {Büttiker}},\
  }\href {\doibase 10.1103/PhysRevB.75.035340} {\bibfield  {journal} {\bibinfo
  {journal} {Phys. Rev. B}\ }\textbf {\bibinfo {volume} {75}},\ \bibinfo
  {pages} {035340} (\bibinfo {year} {2007})}\BibitemShut {NoStop}%
\bibitem [{\citenamefont {Chung}\ \emph {et~al.}(2005)\citenamefont {Chung},
  \citenamefont {Samuelsson},\ and\ \citenamefont
  {Büttiker}}]{chung_visibility_2005}%
  \BibitemOpen
  \bibfield  {author} {\bibinfo {author} {\bibfnamefont {V.}~\bibnamefont
  {Chung}}, \bibinfo {author} {\bibfnamefont {P.}~\bibnamefont {Samuelsson}}, \
  and\ \bibinfo {author} {\bibfnamefont {M.}~\bibnamefont {Büttiker}},\ }\href
  {\doibase 10.1103/PhysRevB.72.125320} {\bibfield  {journal} {\bibinfo
  {journal} {Phys. Rev. B}\ }\textbf {\bibinfo {volume} {72}},\ \bibinfo
  {pages} {125320} (\bibinfo {year} {2005})}\BibitemShut {NoStop}%
\bibitem [{\citenamefont {Parmentier}\ \emph {et~al.}(2012)\citenamefont
  {Parmentier}, \citenamefont {Bocquillon}, \citenamefont {Berroir},
  \citenamefont {Glattli}, \citenamefont {Plaçais}, \citenamefont {Fève},
  \citenamefont {Albert}, \citenamefont {Flindt},\ and\ \citenamefont
  {Büttiker}}]{parmentier_current_2012}%
  \BibitemOpen
  \bibfield  {author} {\bibinfo {author} {\bibfnamefont {F.~D.}\ \bibnamefont
  {Parmentier}}, \bibinfo {author} {\bibfnamefont {E.}~\bibnamefont
  {Bocquillon}}, \bibinfo {author} {\bibfnamefont {J.-M.}\ \bibnamefont
  {Berroir}}, \bibinfo {author} {\bibfnamefont {D.~C.}\ \bibnamefont
  {Glattli}}, \bibinfo {author} {\bibfnamefont {B.}~\bibnamefont {Plaçais}},
  \bibinfo {author} {\bibfnamefont {G.}~\bibnamefont {Fève}}, \bibinfo
  {author} {\bibfnamefont {M.}~\bibnamefont {Albert}}, \bibinfo {author}
  {\bibfnamefont {C.}~\bibnamefont {Flindt}}, \ and\ \bibinfo {author}
  {\bibfnamefont {M.}~\bibnamefont {Büttiker}},\ }\href {\doibase
  10.1103/PhysRevB.85.165438} {\bibfield  {journal} {\bibinfo  {journal} {Phys.
  Rev. B}\ }\textbf {\bibinfo {volume} {85}},\ \bibinfo {pages} {165438}
  (\bibinfo {year} {2012})}\BibitemShut {NoStop}%
\bibitem [{\citenamefont {Hofer}\ and\ \citenamefont
  {Flindt}(2014)}]{hofer_mach-zehnder_2014}%
  \BibitemOpen
  \bibfield  {author} {\bibinfo {author} {\bibfnamefont {P.~P.}\ \bibnamefont
  {Hofer}}\ and\ \bibinfo {author} {\bibfnamefont {C.}~\bibnamefont {Flindt}},\
  }\href {\doibase 10.1103/PhysRevB.90.235416} {\bibfield  {journal} {\bibinfo
  {journal} {Phys. Rev. B}\ }\textbf {\bibinfo {volume} {90}},\ \bibinfo
  {pages} {235416} (\bibinfo {year} {2014})}\BibitemShut {NoStop}%
\bibitem [{\citenamefont {Streltsov}\ \emph {et~al.}(2016)\citenamefont
  {Streltsov}, \citenamefont {Adesso},\ and\ \citenamefont
  {Plenio}}]{streltsov_quantum_2016}%
  \BibitemOpen
  \bibfield  {author} {\bibinfo {author} {\bibfnamefont {A.}~\bibnamefont
  {Streltsov}}, \bibinfo {author} {\bibfnamefont {G.}~\bibnamefont {Adesso}}, \
  and\ \bibinfo {author} {\bibfnamefont {M.~B.}\ \bibnamefont {Plenio}},\
  }\href {http://arxiv.org/abs/1609.02439} {\bibfield  {journal} {\bibinfo
  {journal} {arXiv:1609.02439}\ } (\bibinfo {year} {2016})},\ \bibinfo {note}
  {arXiv: 1609.02439}\BibitemShut {NoStop}%
\end{thebibliography}

%

\end{document}